\documentclass[traditabstract]{aa}

\usepackage[nonamebreak]{natbib}
\usepackage[stable]{footmisc}
\usepackage{amsmath}
\usepackage{txfonts}
\usepackage{placeins}
\bibpunct{(}{)}{;}{a}{}{,}
\usepackage{graphicx}
\usepackage{epstopdf}
\usepackage{ifthen} 
\usepackage[breaklinks, colorlinks, citecolor=blue]{hyperref}
\usepackage{overpic}
\usepackage{fixltx2e}
\usepackage{rotating}
\usepackage{subfigure}
\usepackage{epsfig}
\usepackage{multirow}
\usepackage{array}
\usepackage{color}
\usepackage[usernames,dvipsnames]{xcolor}
\usepackage{psfrag}
\usepackage[T1]{fontenc}
\usepackage{url}
\usepackage{xfrac}
\usepackage{breakurl}
\def\UrlBreaks{\do\ }
\usepackage{xcolor} 
\usepackage{makecell} %

\newcolumntype{L}[1]{>{\raggedright\let\newline\\\arraybackslash\hspace{0pt}}m{#1}}
\newcolumntype{C}[1]{>{\centering\let\newline\\\arraybackslash\hspace{0pt}}m{#1}}

\definecolor{mywarningcolor}{RGB}{208,59,32} 
\newcommand{\warning}[1]{} 

\begin{document}

\topmargin=-1cm
\oddsidemargin=-1cm
\evensidemargin=-1cm
\textwidth=17cm
\textheight=25cm
\raggedbottom
\sloppy

\definecolor{Blue}{rgb}{0.,0.,1.}
\definecolor{LightSkyBlue}{rgb}{0.691,0.827,1.}
\definecolor{Red}{rgb}{1.,0.,0.}
\definecolor{Green}{rgb}{0.,1.,0.}
\definecolor{Try}{rgb}{0.15,0.,1}
\definecolor{Black}{rgb}{0., 0., 0.}

\title{Development and application of metamaterial-based Half-Wave Plates for the NIKA and NIKA2 polarimeters}

\author{G.~Pisano\inst{\ref{Cardiff}}\thanks{ \email{giampaolo.pisano@astro.cf.ac.uk} }
\and A.~Ritacco\inst{\ref{IAS},\ref{LPENS}}
\and A.~Monfardini\inst{\ref{Neel},\ref{GIS}}
\and C.~Tucker\inst{\ref{Cardiff}}
\and P.A.R.~Ade\inst{\ref{Cardiff}}
\and A.~Shitvov\inst{\ref{Cardiff}}
\and A.~Benoit\inst{\ref{Neel},\ref{GIS}}
\and M.~Calvo\inst{\ref{Neel},\ref{GIS}}
\and A.~Catalano\inst{\ref{LPSC},\ref{GIS}}
\and J.~Goupy\inst{\ref{Neel},\ref{GIS}}
\and S.~Leclercq\inst{\ref{IRAM},\ref{GIS}}
\and J.~Macias-Perez\inst{\ref{LPSC},\ref{GIS}}
\and A.~Andrianasolo\inst{\ref{Ipag}}
\and N.~Ponthieu\inst{\ref{Ipag},\ref{GIS}}
}

\institute{
School of Physics and Astronomy, Cardiff University, CF24 3AA Cardiff, UK \label{Cardiff}
\and
Institut d’Astrophysique Spatiale (IAS), CNRS and Université Paris Sud, Orsay, France\label{IAS}
\and
Département de Physique, Ecole Normale Supérieure, 24, rue Lhomond
75005 Paris, France \label{LPENS}
\and
Univ. Grenoble Alpes, CNRS, Grenoble INP, Institut N\'eel, 38042, Grenoble, France\label{Neel}
\and
Groupement d'Interet Scientifique KID, 38000 Grenoble and 38400 Saint Martin d'H\'eres, France\label{GIS}
\and 
CNRS-LPSC, Grenoble, 53 Rue des Martyrs, 38042 Grenoble, France\label{LPSC}
\and 
Institut de Radioastronomie Millimétrique (IRAM), 300 Rue de la piscine, 38406 St-Martin d'Hères, France\label{IRAM}
\and
Univ. Grenoble Alpes, CNRS, IPAG, 38000 Grenoble, France\label{Ipag}
}

\abstract{
\emph{Context.} Large field-of-view imaging/polarimetry instruments operating at millimeter and submm wavelengths are fundamental tools to understand the role of magnetic fields (MF) in channeling filament material into prestellar cores providing a unique insight in the physics of galactic star-forming regions. Among other topics, at extra-galactic scales, polarization observations of AGNs will allow us to constrain the possible physical conditions of the emitting plasma from the jets and/or exploring the physics of dust inside supernova remnants.
The  kilo-pixel NIKA2  camera, installed at the IRAM 30-m telescope, represents today one of the best tools available to the astronomers to produce simultaneous intensity/polarimetry maps over large fields at 260~GHz (1.15 mm).

\emph{Aims.} The polarisation measurement, in NIKA and NIKA2, is achieved by rapidly modulating the total incoming polarisation. This allows in the end to safely isolate the small science signal from the large, un-polarised and strongly variable, atmospheric background. 

\emph{Methods.} The polarisation modulation is achieved by inserting a fast rotating Half-Wave Plate (HWP) in the optical beam. In order to allow wide field-of-view observations, the plate has to be large, with a diameter exceeding 250~mm. The modulation of the polarised signal, at 12~Hz, requires also the waveplate to be sufficiently light. In addition, this key optical element has to exhibit optimal electromagnetic characteristics in terms of transmission and differential phase-shift. For this purpose, three metamaterial HWPs have been developed using the mesh-filter technology. The knowledge acquired in developing the first two single-band HWPs was used to achieve the more challenging performance requirements of the last dual-band HWP. The first and the third waveplates met the requirements for both the NIKA and NIKA2 instruments. 

\emph{Results.} We first illustrate the design, the technical developments, the fabrication and laboratory characterisation of the three mesh-HWPs. The deployment of two such elements in the NIKA and NIKA2 instruments at the 30-meter telescope is then described. We conclude with representative examples of astrophysical maps integrating polarimetry.
}


\authorrunning{}
\titlerunning{Development and application of metamaterial-based HWPs}
\maketitle


\section{Introduction}
\label{sect:introduction}
The polarization state of the radiation coming from astronomical sources carries important information about the astrophysical environments at the origin of the electromagnetic wave. Generally, only a small fraction of the total signal is polarized and this is embedded within a more intense signal arising from unpolarized light and radiation from other astrophysical sources. However, a linearly polarized signal can be modulated and lifted above the noise level by using a rotating Half-Wave Plate (HWP) operating directly on the incoming beam of the instrument. The rotation of the HWP at a frequency $\omega$ induces a rotation of the polarized signal at twice the mechanical angular rotation speed, 2$\omega$. Detectors sensitive to orthogonal polarizations will eventually detect signals at a frequency 4$\omega$ \citep{ritacco2017}.

The New IRAM KIDs Array (NIKA) \citep{Monfardini_2011,catalano2014} instrument, operated at the IRAM 30-meter telescope in the period 2010-2015, has represented a scientific and technological pathfinder for instrumentation based on Kinetic Inductance Detectors (KIDs). The advantage of KIDs, when compared to the competing technology of bolometers, and in the context of polarization modulation, is the fast (i.e. 0.1 $\sim$ ms) response time. NIKA2 \citep{adam2018} is the successor of NIKA and represents, with around 3k-pixels and three distinct arrays, the ultimate instrument for dual-band (150 and 260~GHz) imaging and polarimetry (260~GHz) at the 30-meter telescope. NIKA2 is a general purpose camera open to the astronomical community via periodic competitive calls. The NIKA2 collaboration is, however, independently pursuing five science objectives spanning from cosmology (e.g. observations of clusters of galaxies via the Sunyaev-Zel'dovich effect, deep integration on cosmological fields) to the study of the Interstellar Medium (ISM) in intensity and polarization \citep{Mayet2020,Ruppin2020,Bracco2017,Peretto2020,Lestrade2020}. Among the most innovative science projects is the polarization study 
of large galactic regions, to understand the role of magnetic fields in the 
fragmentation and evolution of star-forming filaments  \citep{arzoumanian2019,andre_2019}. Magnetic fields, still poorly explored, seem to play a crucial role in a large number of different astrophysical processes \citep{crutcher2012}. They can be traced via observations of dust polarization \citep{planckxix}. The NIKA2 polarimeter, and previously the NIKA one, can fill the gap between high-resolution observations provided by interferometers like {\it ALMA, NOEMA} and low-resolution ones as cosmological experiments, e.g. {\it Planck, WMAP} and future LiteBird satellites. For this reason it was important to implement polarimetric capabilities in the instrument concept of NIKA2, taking advantage of the new HWP technology that is described in detail in the present paper. 

At millimeter wavelengths, the quasi-optical modulation of linear polarization can be achieved using different types of HWPs. Diverse solutions have been developed in the past, ranging from birefringent multi-plate designs \citep{a,b,c,d,e,f,g,h}, artificial birefringent material based devices \citep{i,j} and reflective-based solutions \citep{k,l,m}. 

However, in this work we present the novel development of two large bandwidth metamaterial based HWPs, which have been installed and have successfully operated in the NIKA \citep{ritacco2017,ritacco2018} and NIKA2 \citep{adam2018,ritacco2020} instruments. These devices have been developed using the well-established mesh-filter technology \citep{Ade,n}. Previous realizations of mesh-HWPs had either limited bandwidth, of the order of 30\% \citep{o,p}, or non-ideal phase-delays \citep{q}. The goal of this work is to demonstrate high performance across bandwidths of the order of 90\% ($\sim$3:1), never achieved before with this technology.
\section{The NIKA and NIKA2 instruments}\label{sec2}
In this section we briefly describe the NIKA (pathfinder) and NIKA2 (final instrument) optical designs and the requirements for the associated HWPs. We remind that both NIKA and NIKA2 are based on custom dilution cryostats, designed and realized in order to provide the needed base temperature of around 0.1$\sim$K for the KID focal planes (filled arrays). At the same time, a part of the optics is cooled to cryogenics temperatures, contributing thus to reduce to negligible levels the influence of stray radiation.  Both NIKA and NIKA2 are dual-band, mapping simultaneously the same portion of the Sky in the so-called 1mm and 2mm atmospheric windows. NIKA2 has been designed, since the beginning, with the polarisation mapping at 260~GHz in head. On the other hand, NIKA has been adapted \emph{a posteriori} to measure the polarisation, despite in non-optimised conditions, at both 150~GHz and 260~GHz. For more details please refer to paragraphs \ref{nikaoptics} and \ref{nika2optics}.

As already mentioned, the KID detectors allow a fast, i.e. $\geq$10$\sim$Hz, modulation of the incoming polarization. To achieve that, a mechanical rotational frequency of around 3 Hz is applied to the HWP modulator.  The distinct advantage of a fast and continuous modulation is the ability of eliminating almost completely the influence of the atmospheric 1/f noise that is often limiting the sensitivity of the ground-based polarimeters operating at similar wavelengths. 

\subsection{NIKA optics}
\label{nikaoptics}
\begin{figure}
\centering
\includegraphics[width=\columnwidth]{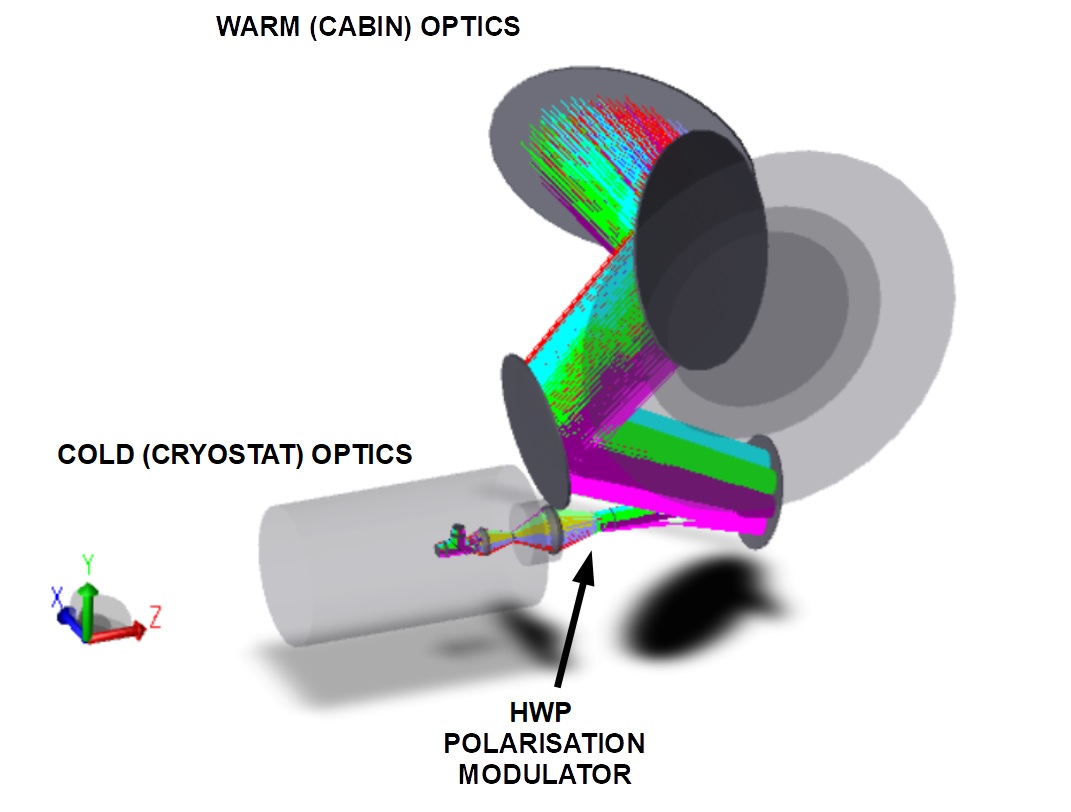}
\caption{NIKA instrument optical layout.}
\label{nika_optical_layout}
\end{figure}

NIKA covered an instantaneous field-of-view of around 2 arc-minutes (diameter), mapped with angular resolution of 18~arc-sec (150~GHz) and 12 arcsec (260~GHz).  The NIKA optics, schematically shown in Fig.~\ref{nika_optical_layout}, is split between a warm part entirely made by reflective elements (cabin optics) and a purely refractive (cryostat optics) section realised using anti-reflection-coated HDPE (High-Density Polyethylene) lenses. NIKA was not intrinsically designed for polarization measurements. For this reason, both the HWP modulator and the fixed wire-grid projection polarizer were located at room temperature, at the intersection of the cold and warm optics sections. This sub-optimal configuration has two unwanted effects: a) since only one polarization is transmitted into the cryostat, the overall quantum efficiency is reduced by a factor of two; b) a non-negligible background is added by the room-temperature wire-grid via reflections of the ambient radiation.  Despite that, the polarization studies in NIKA have produced important science results and most importantly provided crucial lessons for the design of NIKA2. 

The beam size at the HWP location, shown in Fig.~\ref{nika_optical_layout}, is of about 100~mm (diameter).

\subsection{NIKA2 optics}
\label{nika2optics}
\begin{figure}[!ht]
\centering
\includegraphics[width=\columnwidth]{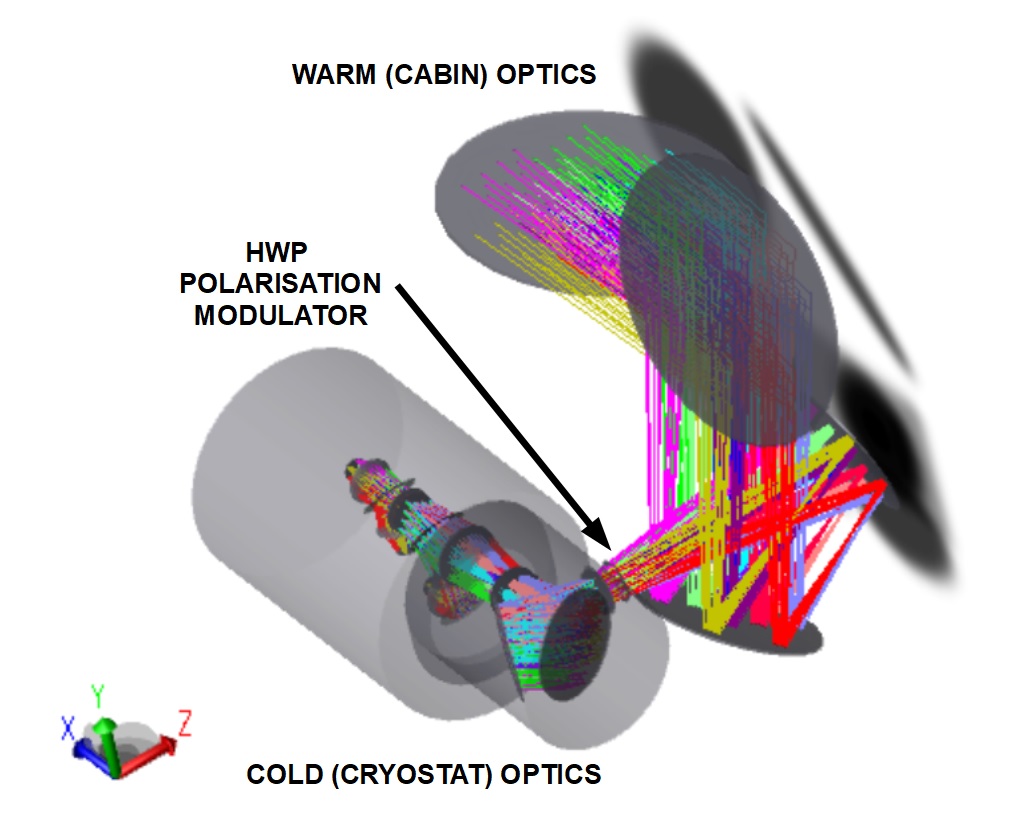}
\caption{NIKA2 instrument optical layout.}
\label{nika2_optical_layout}
\end{figure}

NIKA2 features an instantaneous field-of-view of 6.5~arc-minutes (diameter), mapped with angular resolution (at 260~GHz) of 11~arc-sec. The size of the primary image of the 30-meter telescope, for such a field-of-view, is of the order of 60~cm. This already gives an idea of the dimensions of the optical components that are needed to re-form the image onto the 80~mm (diameter) focal planes (see Fig.~\ref{nika2_optical_layout}). As was the case for NIKA, the NIKA2 optics is split between a warm part entirely made by reflective elements (cabin optics) and a cold section realised mixing mirrors and AR-coated HDPE lenses.  Thanks to its low emissivity, the HWP polarization modulator can sit at room temperature, at the intersection between the two (warm and cold) optics sections. In order to minimize possible systematic, at the HWP location the NIKA2 beam exhibits a pupil; it is in other words completely defocused. The wire-grid projection polarizer is, in NIKA2, mounted at 45~degrees and at base temperature (0.1~K). It feeds two arrays sensitive to orthogonal polarizations, and does not add significant background. For these reasons, we consider that NIKA2 has been designed, since the beginning, also with polarimetric observations in mind.

The size of the beam at the HWP location, shown in Fig.~\ref{nika2_optical_layout}, is of the order of 200~mm in diameter. Please refer to \citep{adam2018} for more details concerning the NIKA2 instrument.  

\begin{figure}[!ht]
\centering
\includegraphics[width=\columnwidth]{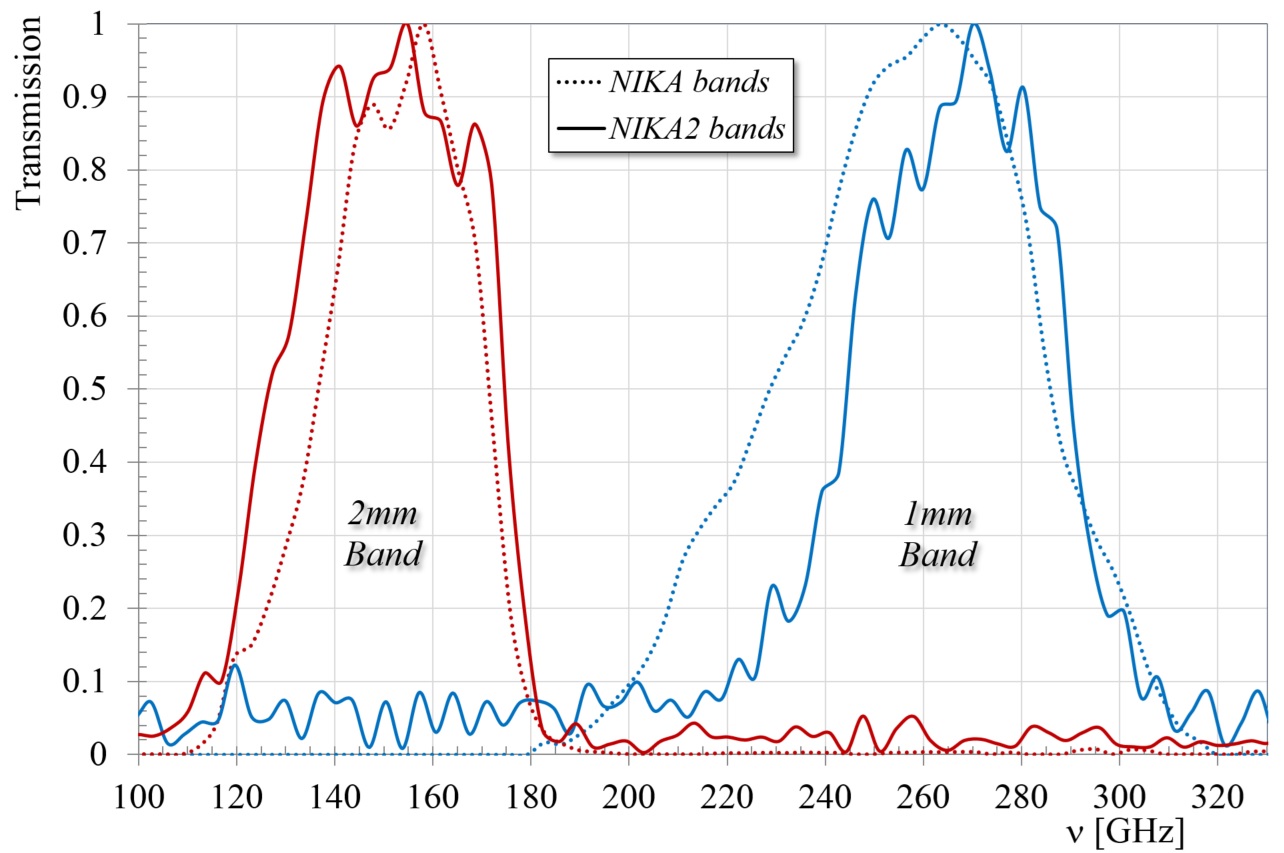}
\caption{NIKA and NIKA2 instruments ``1mm'' and ``2mm'' spectral bands. The different profiles have been taken into account in the design of the different waveplates.}
\label{fig_03}
\end{figure}

As shown in Fig.~\ref{fig_03}, the NIKA and NIKA2 spectral bands are similar, requiring a HWP with good performance between 120 and 300~GHz. 

\section{Mesh HWPs design, manufacture and laboratory characterization}\label{sec3}
\subsection{Mesh-HWPs for NIKA and NIKA2}\label{subsec3.1}
Different approaches, and associated technologies, can be employed to design quasi-optical HWPs operating at millimetre wavelengths \citep{r}. In this work we have pushed forward the development of the mesh-HWP technology, started with the air-gap version \citep{o} which subsequently evolved into more sophisticated dielectrically embedded versions \citep{p,q}. The dual goal was to achieve both a high performance in terms of transmission and a differential phase-shift across large bandwidths (of the order of ~90\%, or 3:1) never achieved before with our technique. 

A single HWP was required to operate simultaneously across the 1mm and 2mm channels of both the NIKA and NIKA2 instruments (Fig.~\ref{fig_03}). 

The work was carried out between the last observational campaigns of the NIKA instrument and the deployment of the new NIKA2 polarimeter at the IRAM telescope. The staggered development phases comprised three mesh-HWPs: one for NIKA (M-HWP$_A$); one for further R\&D (M-HWP$_B$); and one for NIKA2 (M-HWP$_C$). The knowledge and understanding gained during the manufacture and test of these waveplates allowed us to improve the design of the successive plates and to achieve increasingly better performances. Ultimately, this new design has met both the NIKA and NIKA2 instrument requirements.

Before describing the development details of the three HWPs we are going to summarize the design, manufacture and testing procedures which are common for all these devices.
\subsection{Design of the mesh-HWPs}\label{subsec3.2}
\begin{figure}[!ht]
\centering
\includegraphics[width=\columnwidth]{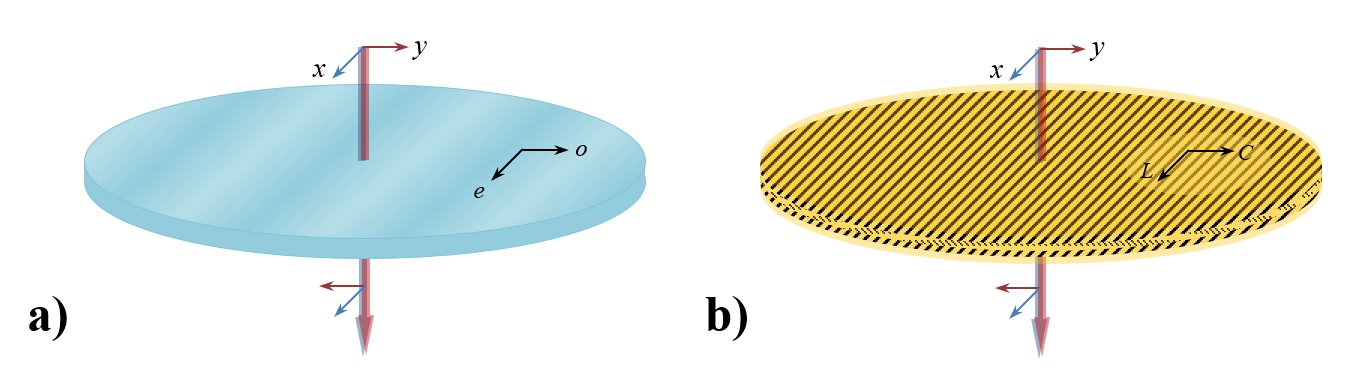}
\caption{ a)Sketch of a birefringent (Sapphire) HWP. The ordinary and extraordinary axes lie across the waveplate. b) Sketch of a dielectrically embedded mesh-HWP. The ordinary and extraordinary axes are replaced by capacitive and inductive axes.}
\label{fig_04}
\end{figure}

The simplest design for a HWP operating at millimetre wavelengths consists of a birefringent plate (usually Sapphire) with its thickness tuned to provide a phase-shift of 180$^{\circ}$ (at a specific frequency) between the ordinary and extraordinary axes (Fig.~\ref{fig_04} a ). 

In order to achieve large bandwidths it is possible to stack many plates together with different orientations following the well-known Pancharatnam designs \citep{a}. The relatively large refractive index of the Sapphire plates (\textit{n}$\sim$3.1) needs to be matched to free-space by means of multi-layer anti-reflection coatings (ARCs) on both sides of the plate. The difficulties associated with the development of ARCs operating at cryogenic temperatures due to differential thermal contraction effects, the fragility of the device and the overall mass of the HWP pushed us to investigate alternative solutions based on the mesh-filter technology.

\begin{figure}[!ht]
\centering
\includegraphics[width=\columnwidth]{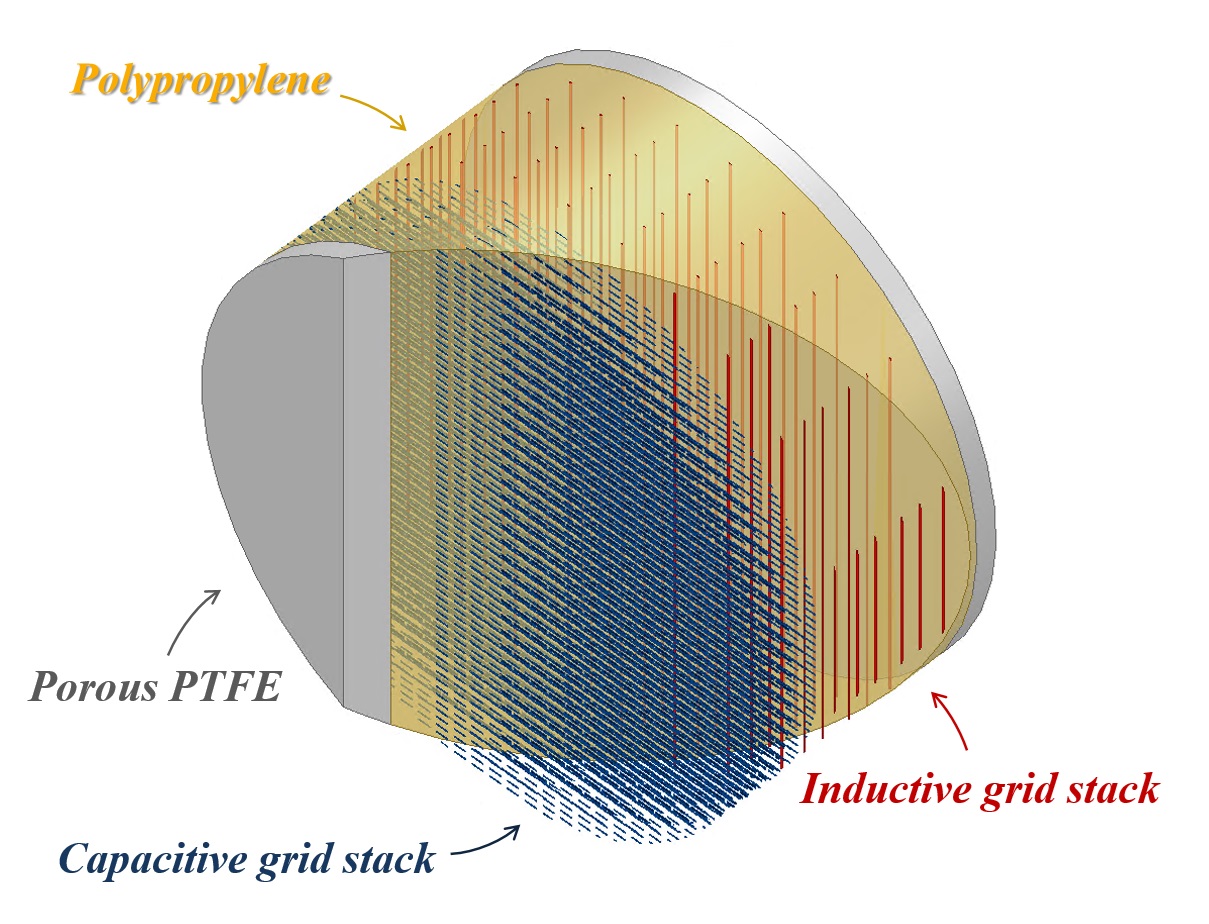}
\caption{CAD design example of a dielectrically embedded mesh-HWP. The cylindrical section drawn here has a diameter of 6mm.}
\label{fig_05}
\end{figure}

A mesh-HWP is a combination of two anisotropic mesh-filters interacting differently (and separately) with two orthogonal polarizations: a capacitive low-pass filter along one axis and an inductive high-pass filter along the orthogonal axis, both providing phase-shifts in opposite directions (Fig.~\ref{fig_04}b). These filter elements can be realized with thin copper grids embedded within polypropylene (n$\sim$1.5). They must show high in-band transmission while keeping an almost frequency-independent differential phase-shift, close to 180$^\circ$. In order to operate across large bandwidths ($\sim$90\%) both the capacitive and the inductive filters should comprise at least six grids. The inductive grids resemble parallel lines whereas the capacitive grids look like parallel dashed-lines (see Fig.~\ref{fig_05}). The final mesh-HWP consists of a capacitive stack in series with an inductive stack but oriented in orthogonal directions. Ideally each stack should be transparent to in-band radiation in its orthogonal direction so as not to deteriorate the performance of the other stack. Finally, single layers of porous-PTFE
(n$\sim$1.25) applied on both sides of the plate can be used as efficient anti-reflection coatings.

The mesh-HWP design is carried out by means of a specifically developed propagation matrix code. The grids are treated like lumped elements in a transmission line circuit and their admittances are accurately calculated using finite-element analysis (Ansys HFSS \url{www.ansys.com}) along both polarization directions. Optimisation procedures are implemented to obtain high transmission and the required differential phase-shift within the required frequency ranges. The non-ideal behaviour of the grids, the multiple reflections within them, and the interactions between the stacks are inherently taken into account within the code. Details of these procedures have been reported elsewhere \citep{p}.
\subsection{Manufacture of the mesh-HWPs}
\label{subsec3.3}

In summary, the design of the metal-mesh HWP involves a stack of capacitive low-pass elements (C) and a stack of inductive high pass elements (L), bonded together. Each stack consists of six grids, different in each of the three designs presented, aligned and stacked in a symmetrical arrangement, as shown in Fig.~\ref{fig_05}. Layers of  anti-reflection coating (ARC) are eventually applied on both sides of the device.      

The grids are made by photolithographically patterning high purity 400nm thick copper layers, which have been thermally evaporated onto polypropylene substrates.  The lithography and wet-chemical etching are highly controlled, producing uniform, replicable patterns.  However, to further improve device performance, the geometry of each grid is measured and  fed back into the model code in order to optimize the device at the assembly stage. 

Alignment of the grids over a large area is achieved with the aid of fiducial markers and back-lighting and takes the expertise of an experienced cleanroom process technician. Once aligned the individual C and L stacks are hot-bonded together using a standard oven with a highly controlled pressure-temperature cycle over 3 days. These separate stacks are also spectrally characterised, using the setup described in section \ref{subsec3.4}, and a further optimisation process performed.  Then the C and L stacks are orthogonally aligned and hot-bonded.  Finally, an ARC layer of porous-PTFE is applied to both surfaces and also hot-bonded in place.  
\subsection{Experimental setup for laboratory tests of the mesh-HWPs}\label{subsec3.4}
The NIKA/NIKA2 HWP elements and final devices were all tested with a Martin-Puplett polarizing Fourier Transform Spectrometer (FTS).  The FTS was configured for optimized performance in the 100 to 330 GHz range, using photolithographically-etched polarizers. The FTS output was coupled to a pumped 4He germanium bolometric detector allowing high sensitivity measurements in this FIR spectral regime.  Data were taken at room temperature with 120-interferogram averages recorded and Fourier-transformed; these data are ratioed against background spectra from the FTS with no sample installed. 
 
\begin{figure}[!ht]
\centering
\includegraphics[width=\columnwidth]{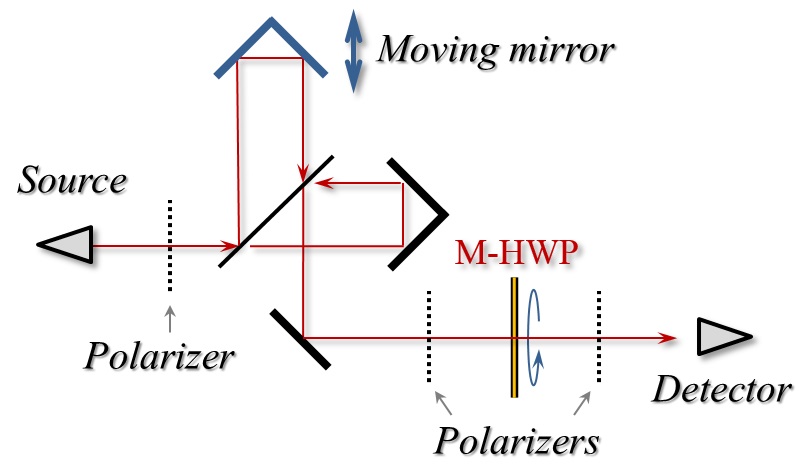}
\caption{Martin-Puplett Fourier Transform Spectrometer setup used for the characterization of the mesh-HWPs.}
\label{fig_06}
\end{figure}

 The polarisation in the FTS was defined by a polarizer in front of the source. Two additional polarizers aligned with the first one were located at the output the FTS. The HWPs under test were positioned in between the output polarizers and could rotate around their optical axis (see Fig.~\ref{fig_06}).  Three sets of measurements were performed for each HWP by aligning their C- and L-axis with the polarizers and at 45$^{\circ}$ from these.

\begin{figure}[!ht]
\centering
\includegraphics[width=\columnwidth]{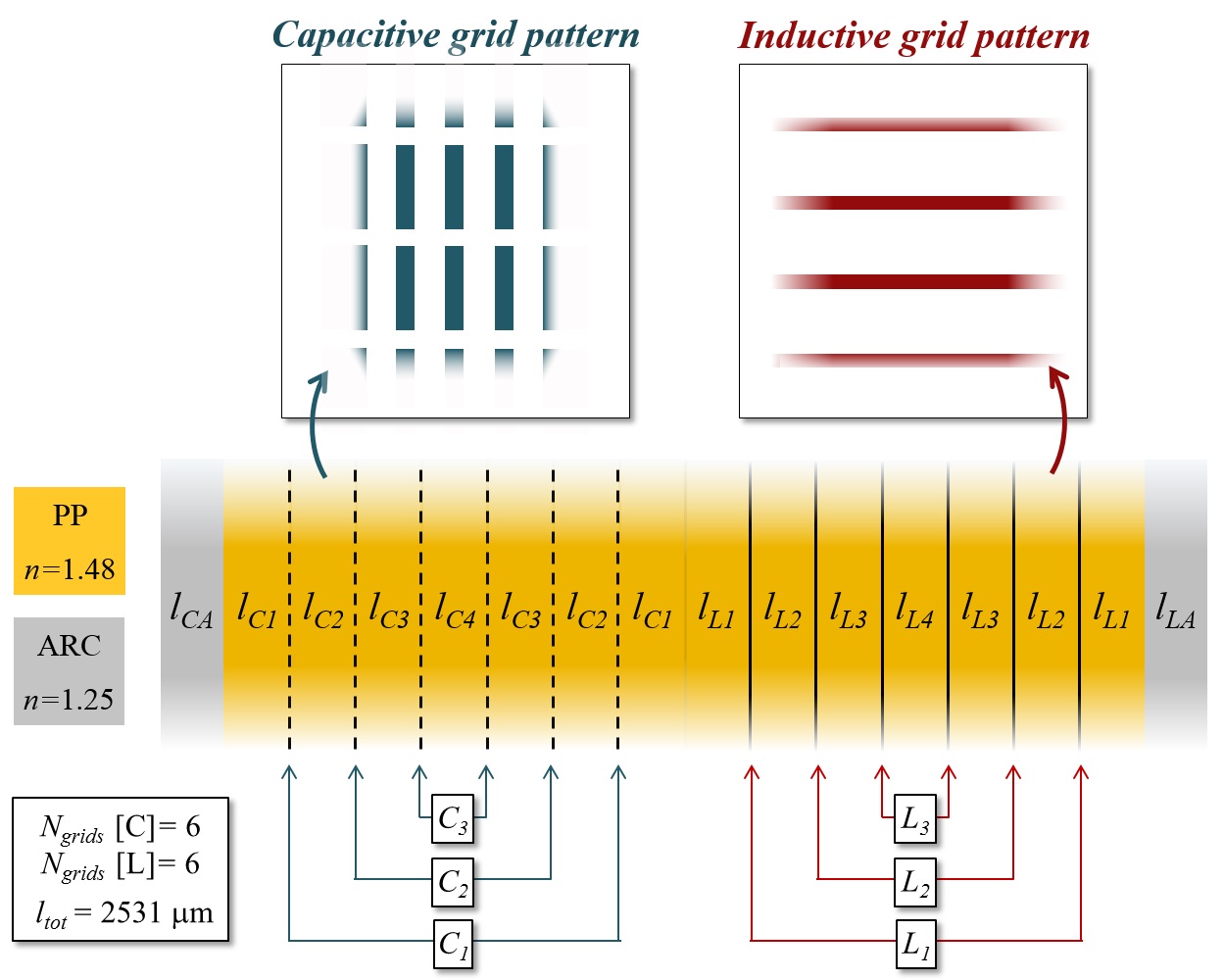}
\caption{M-HWP$_A$ design details: sketch of the capacitive and inductive metal grid geometries and their symmetric arrangement. This device consists of 6 capacitive and 6 inductive grids embedded into polypropylene, with one layer of AR-coating bonded on its sides. The overall thickness is $\sim$2.5 mm.}
\label{fig_07}
\end{figure}

\section{NIKA mesh-HWP first prototype (M-HWP$_A$)}\label{sec4}
The target performance of the NIKA mesh-HWP was a transmission along the capacitive and inductive axes above 90\%, i.e. T$_C$ $\geq$ 0.90 and T$_L$ $\geq$ 0.90, and a differential phase-shift between the axes $\Delta\Phi_{C-L}$= 180$^\circ$ $\pm$ 20$^\circ$ averaged within the 1-mm band. These latter values would guarantee an averaged polarization modulation efficiency around 97\%. Good performance of the HWP within the 2-mm band was desirable but not a requirement. However, as it will be discussed in Sec.\ref{sec7}, polarization measurements on astronomical sources have been eventually performed in both bands.

The optimization procedure of the propagation matrix code was set to minimize the phase-shift error and maximize the transmissions across the whole range of frequencies including the two bands, i.e. from 110 to 320 GHz. The result of this optimisation provided a 12-grid device, six capacitive and six inductive grids, with the geometries and the spacing reported in Fig.~\ref{fig_07}. The six inductive grids were all identical whereas three types of capacitive grids were required. All the evaporated grids had a copper thickness of 400nm. Single-layer anti-reflection coatings were added on both sides of the waveplate to ensure good matching to free space. The conductive losses of the copper grids and the dielectric losses of the polypropylene substrates were taken into account in the simulations. The polypropylene loss-tangent was extracted from previous measurements on planar slabs of material.

\begin{figure}[!ht]
\centering
\includegraphics[width=\columnwidth]{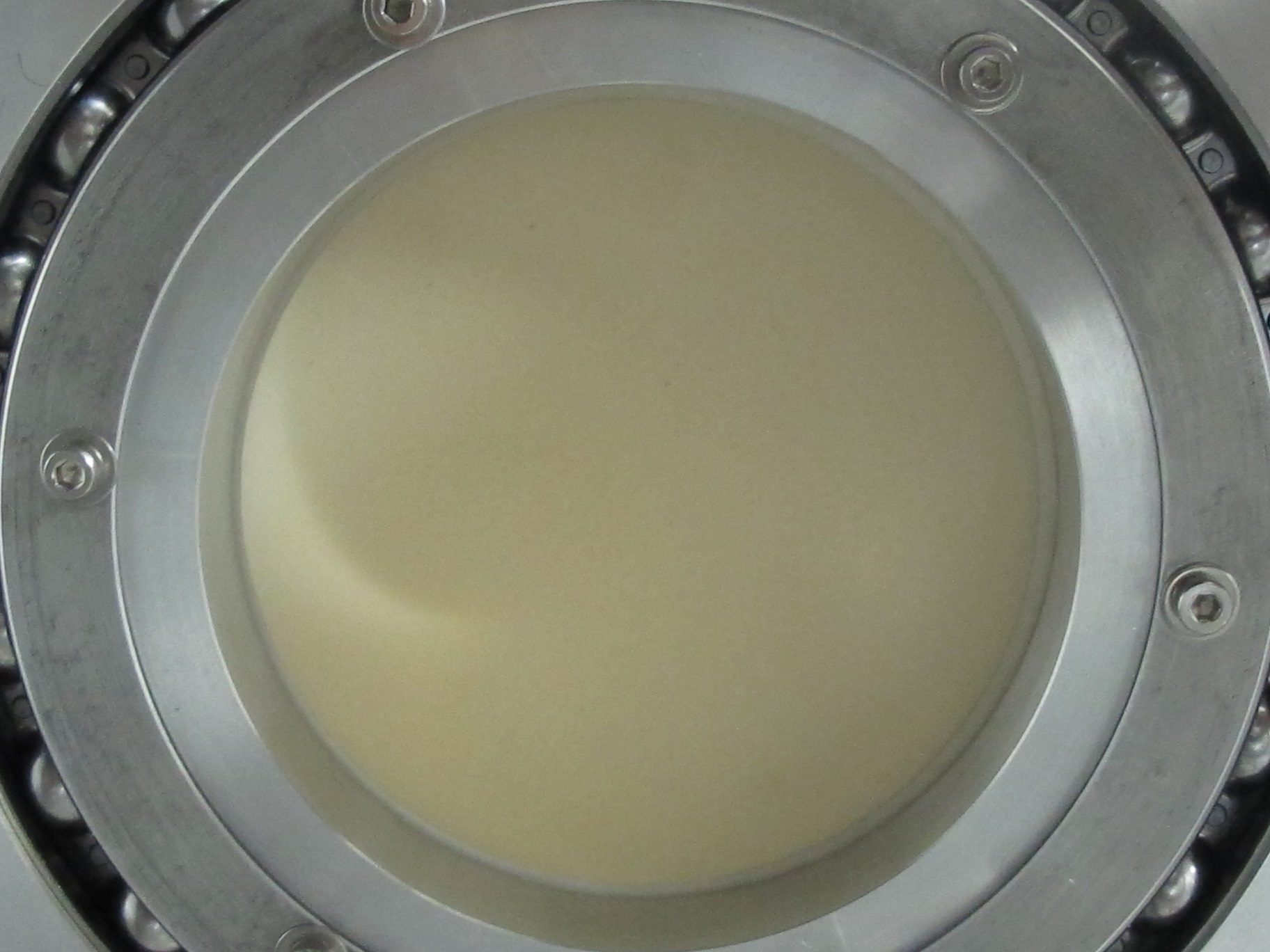}
\caption{The NIKA M-HWP$_A$ first prototype.}
\label{fig_08}
\end{figure}

The twelve grids were manufactured, the capacitive and inductive stacks assembled, hot-pressed and AR-coated following the procedures described in Sec.~\ref{subsec3.3}. A picture of the M-HWP$_A$ device, during its operation in NIKA instrument, is shown in Fig.~\ref{fig_08}. The device had a diameter of 20 cm and a thickness of $\sim$2.5 mm. 

\begin{figure}[!ht]
\centering
\includegraphics[width=\columnwidth]{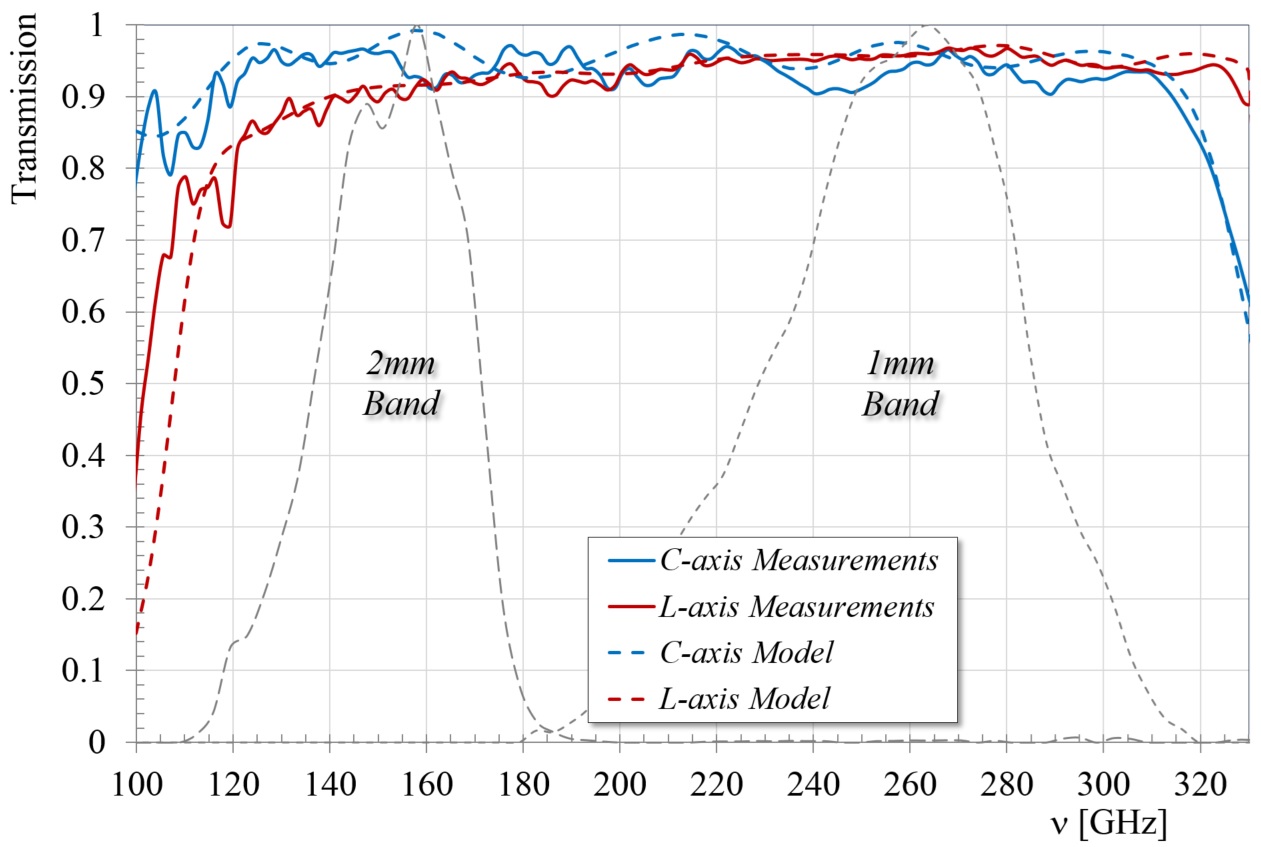}
\caption{M-HWP$_A$: Capacitive and inductive axes modelled and measured transmissions.}
\label{fig_09}
\end{figure}

\begin{figure}[!ht]
\centering
\includegraphics[width=\columnwidth]{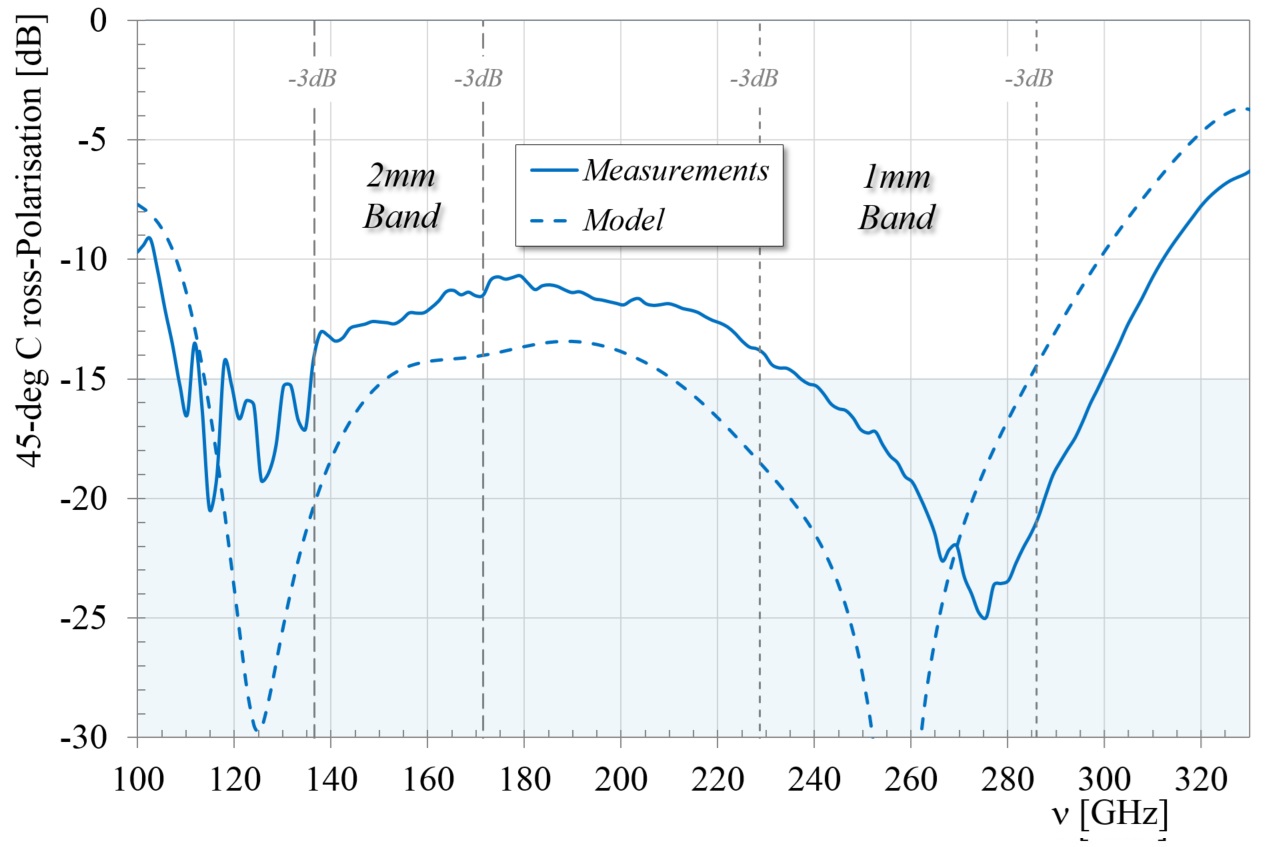}
\caption{M-HWP$_A$: 45$^\circ$ cross-polarization model and measurements.}
\label{fig_10}
\end{figure}

\begin{figure}[!ht]
\centering
\includegraphics[width=\columnwidth]{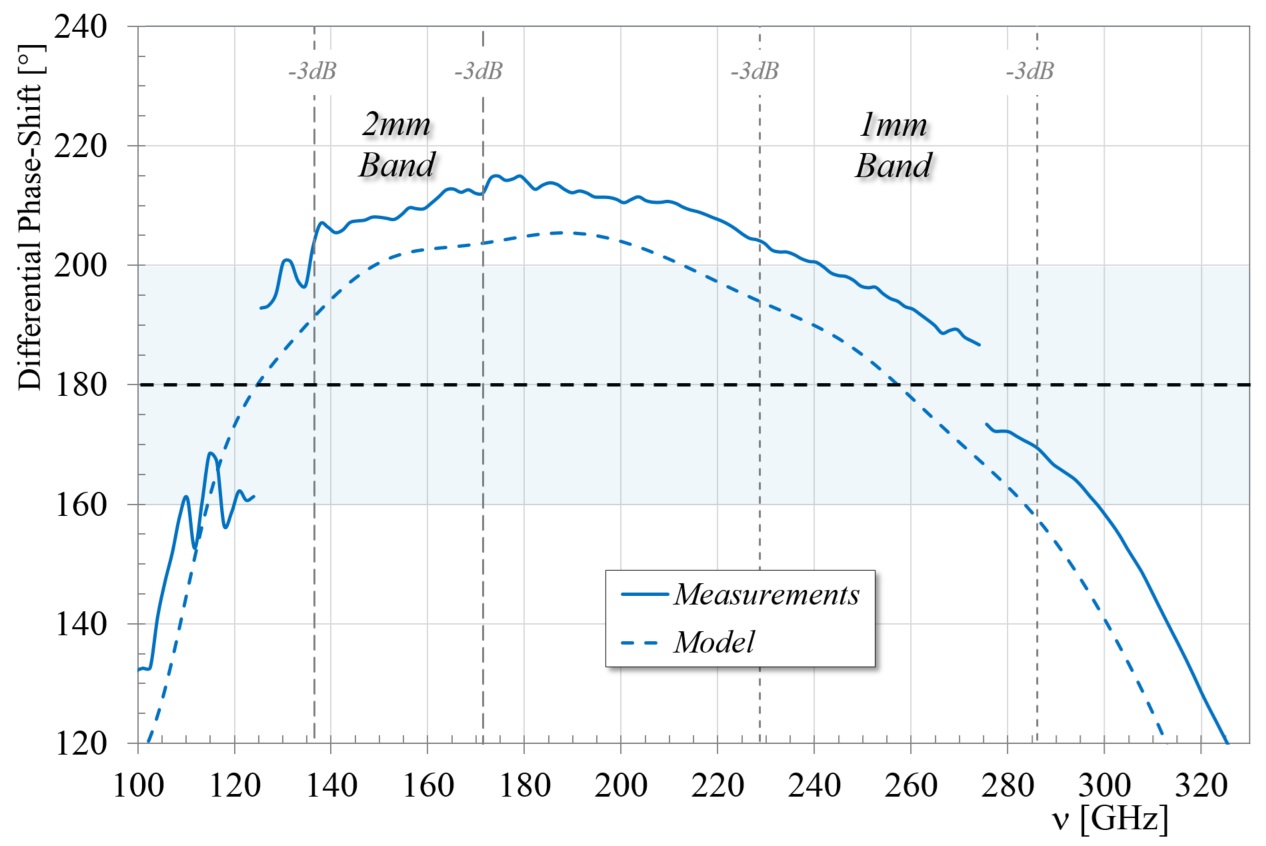}
\caption{M-HWP$_A$: Differential phase-shift model and measurements.}
\label{fig_11}
\end{figure}

Laboratory FTS measurements were performed using the setup discussed in Sec.~\ref{subsec3.4}. The transmission measurements along both axes and the 45$^\circ$ cross-polarization are reported, together with the model expectations in Fig.~\ref{fig_09} and Fig.~\ref{fig_10}. The differential phase-shift can be extracted by combining the previous three measurements, as discussed elsewhere \citep{m}. The model and the results are reported in Fig.~\ref{fig_11}. The phase-shift jump around 180$^\circ$ is due to the sign indeterminacy of the arc-cos function used to extract these values and to the FTS noise, creating a systematic offset from 180$^\circ$.

Although the resulting transmissions can be seen to be above 90\% across the whole 1mm band, there is some disagreement between measurement and simulation. The expected ripples in the capacitive axis transmission are slightly out-of-phase with the measurements. The low-frequency cut-off of the inductive stack is slightly shifted compared to the model (Fig.~\ref{fig_09}). The 45$^\circ$ cross-polarization is a little bit higher than expected, implying a systematic deviation in the differential phase-shift (Fig.~\ref{fig_10} and Fig.~\ref{fig_11}). 

We want to note that the 45$^\circ$ cross-polarization, which is a measure of the differential phase-shift error, is different from the cross-polarization leakage. The latter is the amount of signal leaking into the orthogonal direction when the signal polarization is aligned with one of the HWP axes. In mesh-HWPs this signal is theoretically null due to the symmetry of the design. However, small inaccuracies in the alignment of the grids during the manufacturing processes (at the level of ~$\pm$1$^\circ$) can lead to the appearance of cross-polarization leakages below the -30 dB level, which are much lower than the averaged 45$^\circ$ cross-polarization measurements [MHWP]. 

\begin{figure}[!ht]
\centering
\includegraphics[width=\columnwidth]{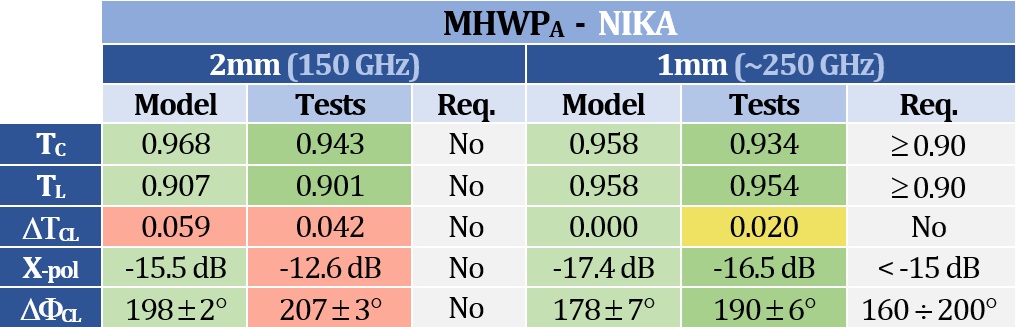}
\caption{Originally modelled, measured and required performance parameters of the M-HWP$_A$ across the two NIKA bands. The parameters are: the axes transmissions T$_C$ and T$_L$; the differential transmission $\Delta$T$_{CL}$; the cross-polarization X-pol; and the differential phase-shift between the axes $\Delta \Phi_{CL}$. The values are averaged and weighted using the NIKA band profiles. The background colours indicate: non-ideal performance (red); acceptable (yellow); and good performance (green). Note that this waveplate was designed to work only in the NIKA 1-mm band.}
\label{tab_1}
\end{figure}

The M-HWP$_A$ transmission along the axes and the differential phase-shifts integrated and weighted across the 1-mm and 2-mm bands are reported in the data in Fig.\ref{tab_1}. The performance of this first device met all the NIKA requirements, in terms of transmission and differential phase-shift. It was installed in the NIKA instrument and successfully used for astronomical polarimetry observations, as discussed in Sec.~ \ref{sec7}. 

Although the device performed well within the 1-mm band, we wanted to investigate the origin of the discrepancies described earlier in order to predict them and improve the performance of the following prototypes. The roots of these deviations lie in the assumptions that we have made in our models, which were based on previously manufactured mesh-devices. For example, the models assumed the refractive index of the polypropylene to be \textit{n}$_{PP}$ = 1.48 and an electrical conductivity of the copper grids being $\sigma_{Cu}$ = 4e7 S m$^{-1}$. The validity of these values, as well as those of other geometrical parameters, will be questioned and investigated in detail in the next two sections. We note that M-HWP$_A$ had also a 2\% unbalance between the axes transmission within the 1-mm band, although it differed only by 1\% across a 98\% bandwidth (110-320 GHz). This was due to the ripples of the capacitive axis transmission (in the 1-mm band) being out–of-phase with respect to the model (see Fig.~\ref{fig_09}). This kind of deviation will be also addressed in the design of the following prototypes.

\section{NIKA mesh-HWP second prototype (M-HWP$_B$)}\label{sec5}
Given the results achieved with the first prototype we manufactured the second waveplate, adopting a more rigorous step-by-step approach. We produced new grids using the same masks of the MHWP$_A$ device and took microscope pictures and measurements of the geometrical details of them all. The period of the grids was found to be as expected, within the measurement error, i.e. below 1$\mu$m. However, we noticed a general over-etching effect associated with the photolithographic processes. The widths and the lengths of the copper lines were systematically reduced by a few microns. The capacitive dash-lines were narrower and shorter respectively by $\sim$3\% and $\sim$6\%. The effect on the inductive lines was much larger given their nominal width of 9 $\mu$m and their actual realizations of $\sim$7 $\mu$m (22\% narrower). Such a large relative change implied not only having less effective grids but also higher losses, due to the thin layer of copper (400nm). We note that, although similar effects were known in the production of mesh filters, in our case the ratio between the period of the grids and the line widths was much larger. This implied removing higher percentages of copper and different etching procedures.

The above effects implied that the grids were less reactive and so introduced individual phase-shifts smaller than expected. The impact on the capacitive-grid stack was limited to the shift in frequency of the transmission ripples seen in MHWP$_A$, whereas the effect on the inductive-stack was more relevant because it implied the addition of higher losses due to their higher resistance. 

 In order to accurately take into account the above effects, each grid was simulated again with finite-element analysis using the actual measured values. The new and more realistic admittances were extracted and inserted in the propagation matrix code. A new optimisation was run to re-design the device with the appropriate corrected spacing between the 6-L and 6-C grids. However, the required performance in terms of phase-shift could not be achieved anymore, even by rearranging the grids, due to the reduced efficiency inductive grids. The differential phase-shift curve moved up by 20 degrees and its averaged values within the band were outside the requirement range.
 
 There were two options to proceed further: i) produce new masks with wider lines to take into account the over-etching effects; ii) add another inductive grid to the current design to compensate for the phase deficit. We decided to proceed with the latter option because this was an R\&D phase and the device was not meant to be used in any instrument requiring high performance across both bands. What was required at that stage was to develop a technique that would allow us to reproduce the experimental data more accurately, regardless of the actual performance. 

\begin{figure}[!ht]
\centering
\includegraphics[width=\columnwidth]{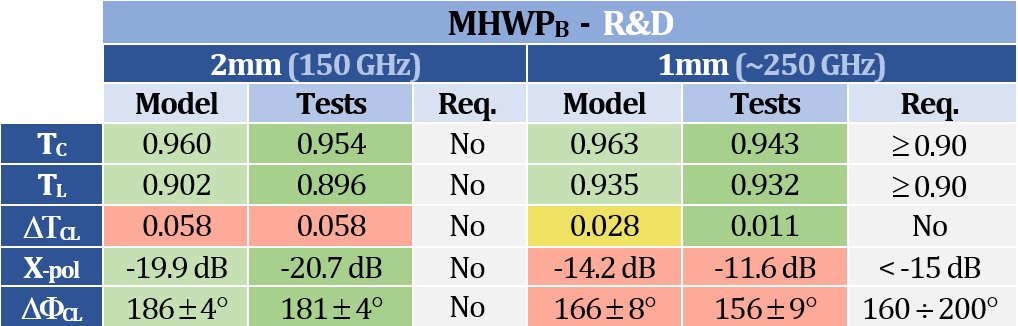}
\caption{Originally modelled, measured and required performance parameters of the M-HWP$_B$ across the two NIKA bands. The labels and colours are the same as described in the data in Fig.~ \ref{tab_1}. Note that this plate was part of an R\&D phase and not meant to work either in the NIKA or NIKA2 instrument.}
\label{tab_2}
\end{figure}

The new 6-C and 7-L recipe was similar to the one sketched in Fig.~\ref{fig_07} but slightly thicker, $\sim$ 3.3 mm. The inductive stack was made of these lossy grids and the performance within the 2-mm band was expected to present a large unbalance in transmission between the two axes (of the order of 6\%). The MHWP$_B$ modelled performance parameters, integrated and weighted within the NIKA bands, are reported in the data in Fig.~ \ref{tab_2}. 

The new capacitive and inductive stacks were assembled and hot-pressed as previously. This time, before the final assembly we wanted to check and test each stack individually. The hot-pressed capacitive and inductive stacks were thicker than expected by 9$\mu$m (0.7\%) and 5$\mu$m (0.3\%) respectively. These values are consistent with the tolerances associated with the large number of polypropylene films used for stacks of these thicknesses. 

\begin{figure}[!ht]
\centering
\includegraphics[width=\columnwidth]{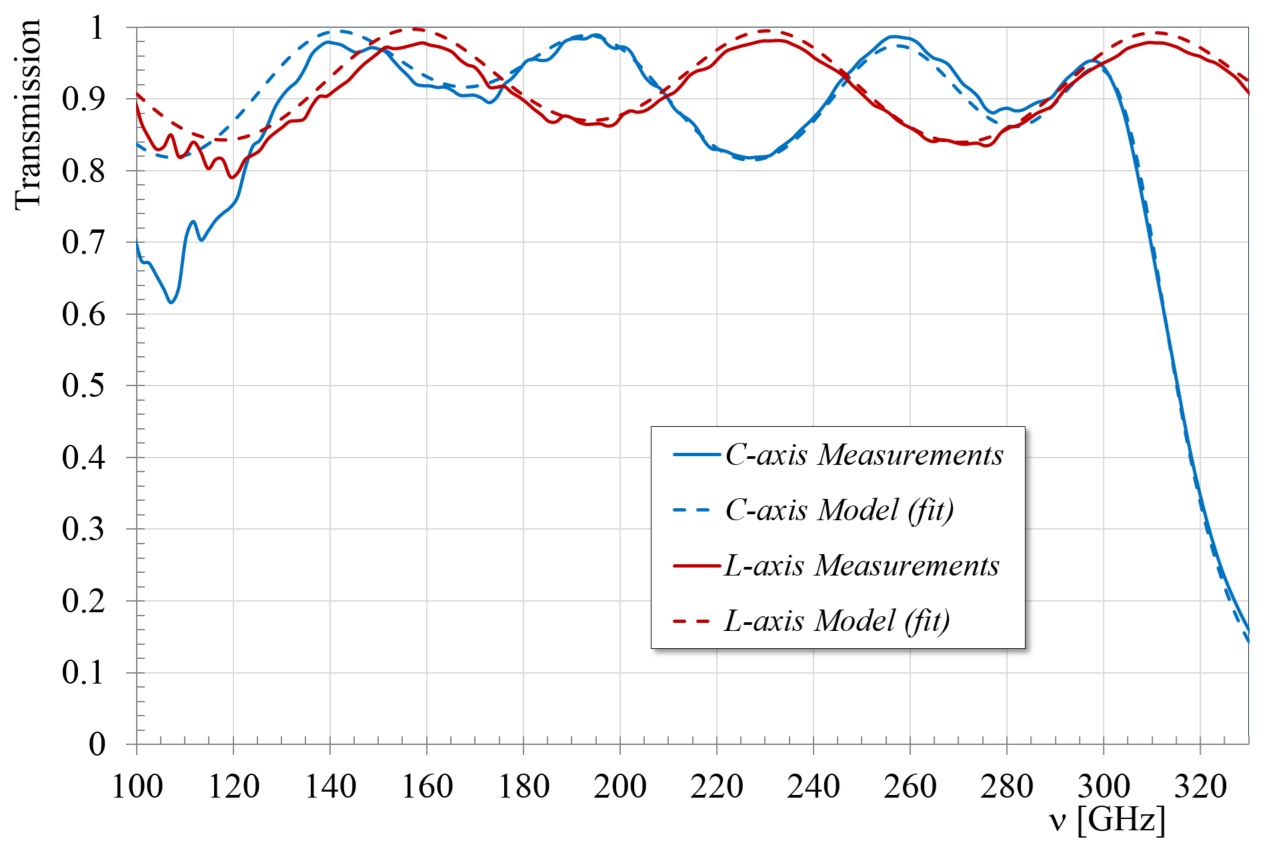}
\caption{M-HWP$_B$: Capacitive stack fitted model and measured transmissions along the reactive and transparent axes. The reactive axis (C) is essentially a low-pass filter while the transparent one (L) is similar to a dielectric slab.}
\label{fig_12}
\end{figure}

\begin{figure}[!ht]
\centering
\includegraphics[width=\columnwidth]{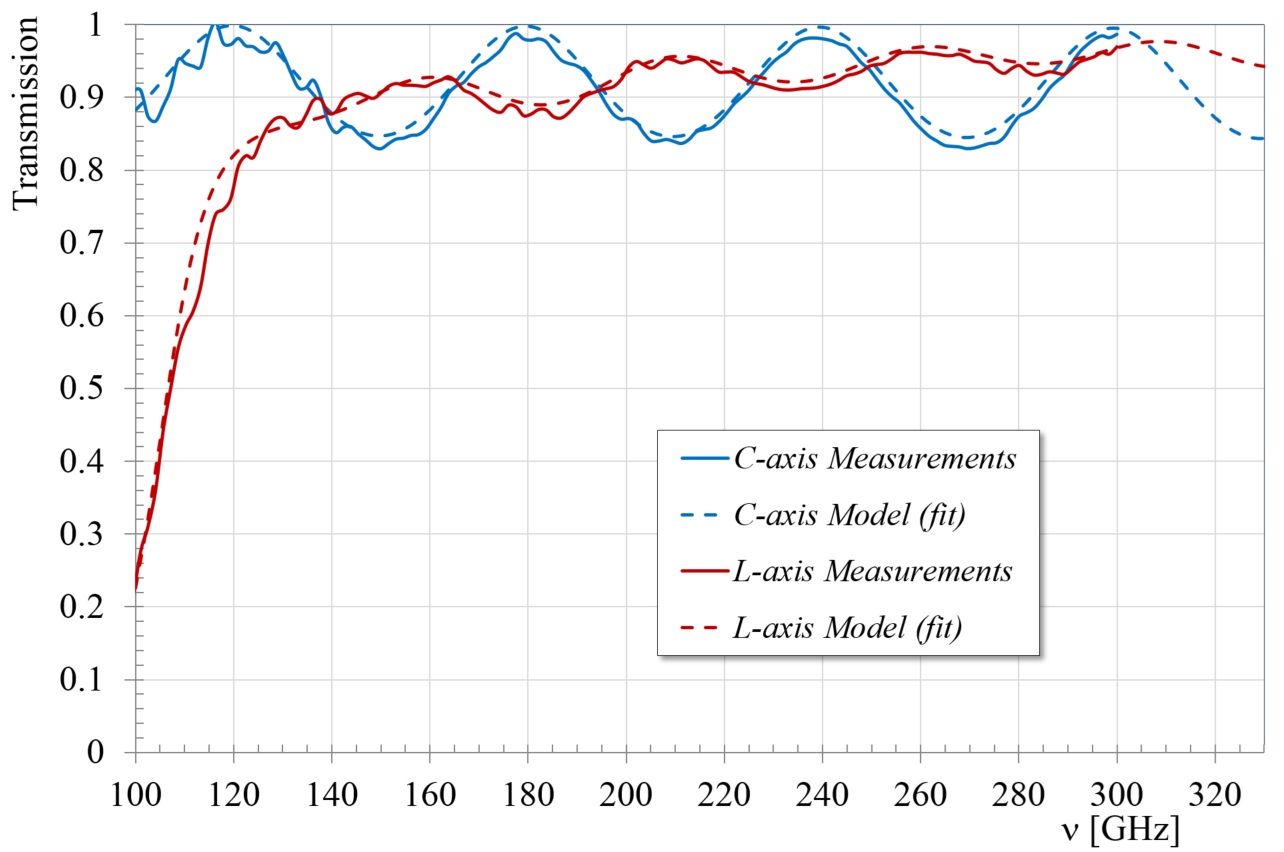}
\caption{M-HWP$_B$: Inductive stack fitted model and measured transmissions along the reactive and transparent axes. The reactive axis (L) looks like a high-pass filter for polarization aligned with it, while the transparent axis (C) like a dielectric slab.}
\label{fig_13}
\end{figure}

Both stacks were tested with the FTS. The measured transmissions along the reactive and transparent axes of each stack are reported in Fig.~\ref{fig_12} and Fig.~\ref{fig_13}. In order to achieve agreement between the model and the measured data, an optimization was run by varying the only two parameters left that could still be different from the predictions: the refractive index of the polypropylene and the electrical conductivity of the copper grids. The best fit was achieved when these parameters were respectively \textit{n}$_{PP}$ $\simeq$ 1.512 and $\sigma_{Cu}$ = 3e7 S m$^{-1}$. The refractive index increase, from the nominal value 1.48, is believed to be a consequence of the hot-pressing process of the many thin polypropylene layers (of the order of 200) which then became a single plastic block. In fact, the bi-axially oriented polypropylene substrate material is known to contract during the hot-pressing process; whilst this is controlled to a certain extent, there is inevitably some slight increase in the material density over progressive pressings. The fitted copper conductivity is consistent with values expected by evaporation processes and measured previously with some mesh-filter devices. We note in Fig.~\ref{fig_12} and Fig.~\ref{fig_13} that the reactive C- and L-axis show the expected low-pass and high-pass cut-off while the transparent axes behave similarly to dielectric slabs.


\begin{figure}[!ht]
\centering
\includegraphics[width=\columnwidth]{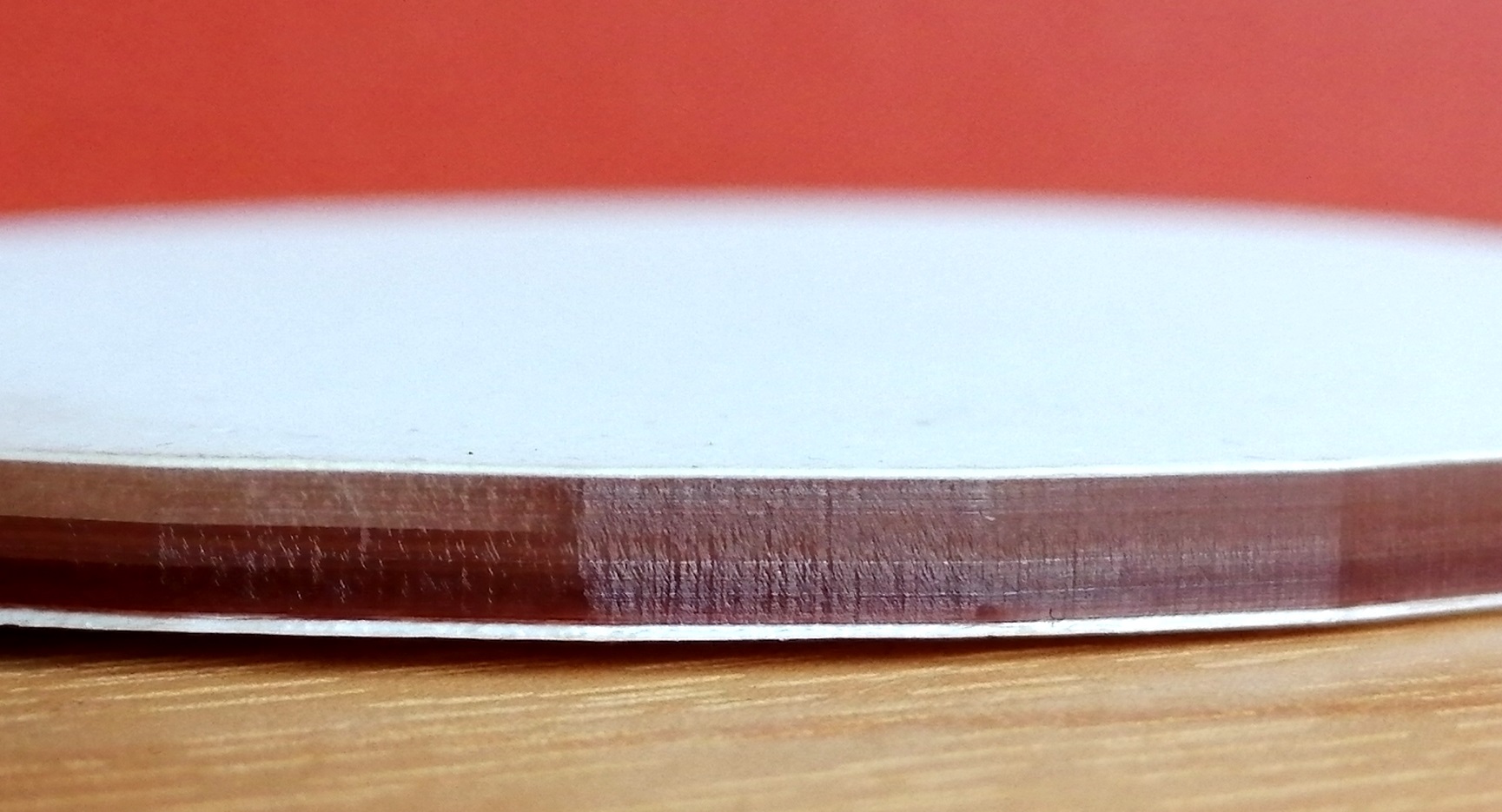}
\caption{Cross-section details of the second prototype M-HWP$_B$. The white layers are the porous-PTFE anti-reflection coatings while the brown part is the polypropylene substrate hosting all the embedded copper grids.}
\label{fig_14}
\end{figure}

\begin{figure}[!ht]
\centering
\includegraphics[width=\columnwidth]{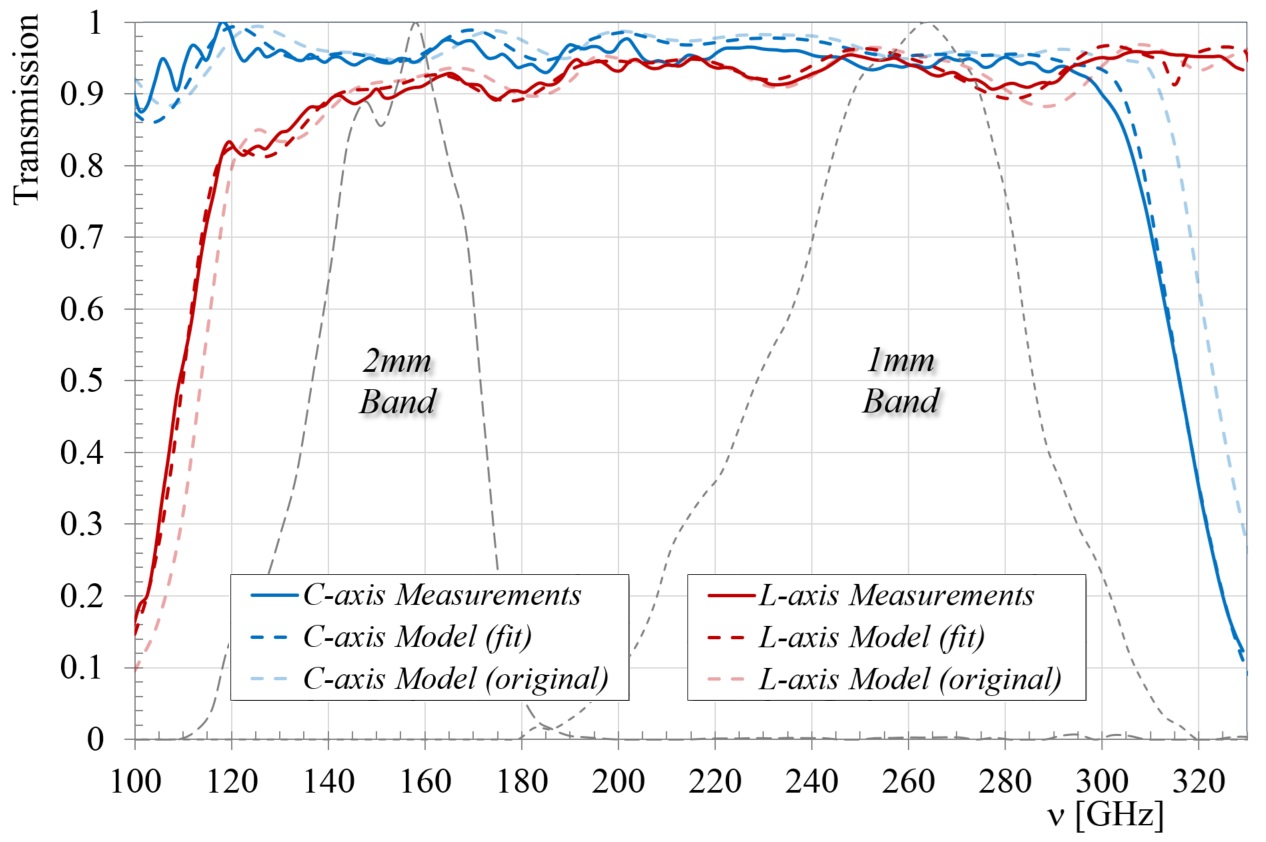}
\caption{M-HWP$_B$: Capacitive and inductive axes modelled and measured transmissions.}
\label{fig_15}
\end{figure}

\begin{figure}[!ht]
\centering
\includegraphics[width=\columnwidth]{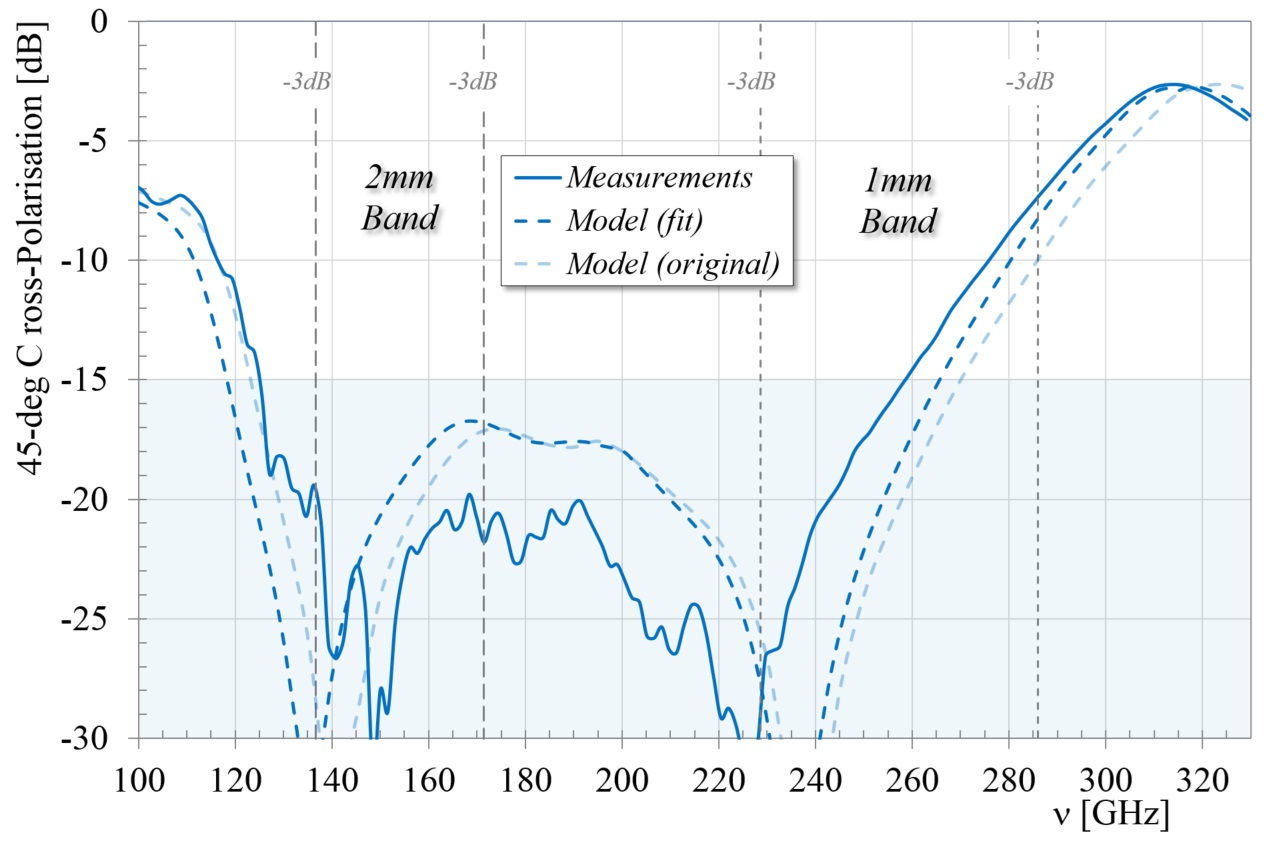}
\caption{M-HWP$_B$: 45$^{\circ}$ cross-polarization model results and measurements.}
\label{fig_16}
\end{figure}

\begin{figure}[!ht]
\centering
\includegraphics[width=\columnwidth]{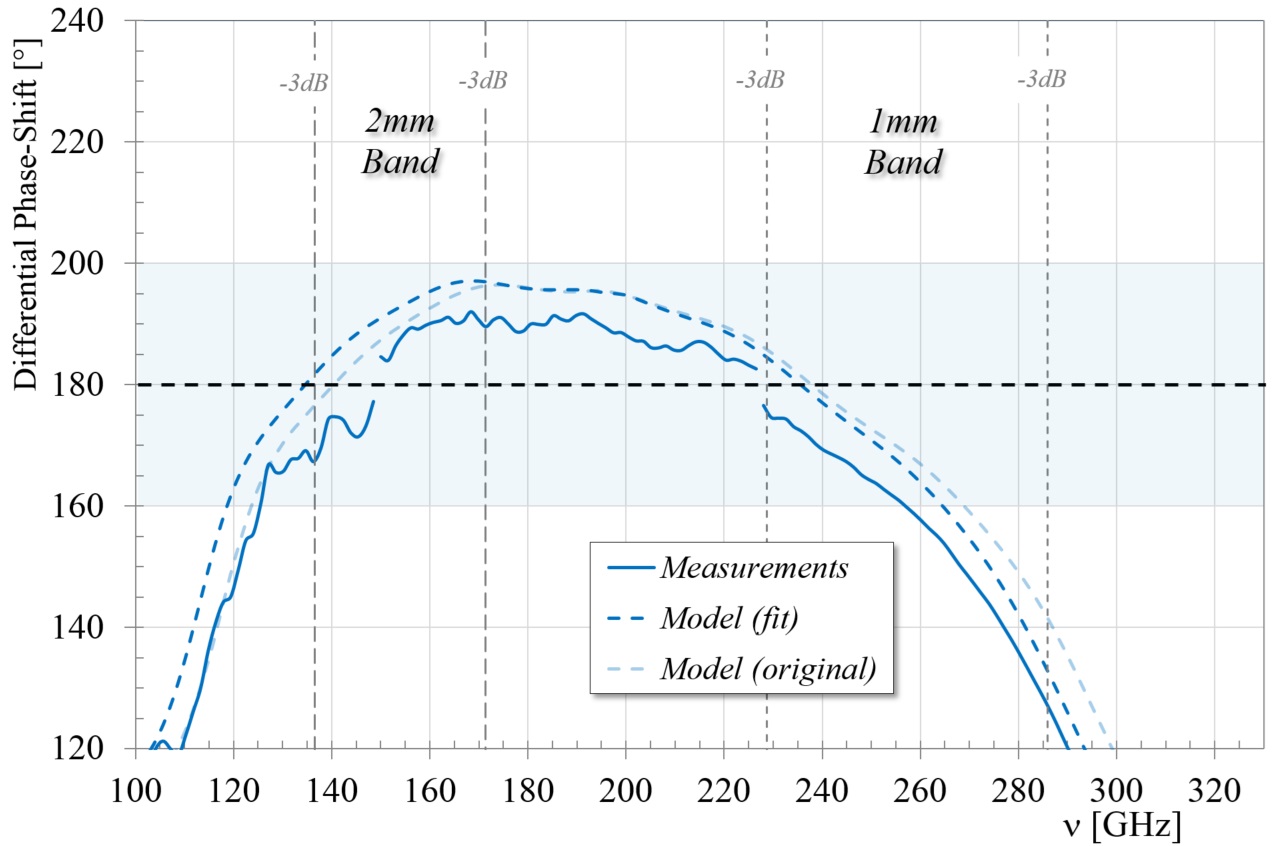}
\caption{M-HWP$_B$: Differential phase-shift model results and measurements.}
\label{fig_17}
\end{figure}

Given the good agreement between data and fitted models, the two stacks were eventually bonded together and AR-coated. The final device had an overall thickness of $\sim$3.3 mm. Fig.~\ref{fig_14} is a picture with details of this second prototype. The expected performance of the original design and the test results of the M-HWP$_B$ are shown in Fig.~\ref{fig_15}, Fig.~\ref{fig_16} and Fig.~\ref{fig_17}. In order to better fit the experimental data, we left the refractive index to be a variable parameter which eventually transpired to be higher than those measured for the individual stacks, i.e. \textit{n}$_{PP}$=1.535. This is attributable again to the further and final hot-bonding process. The model results obtained using this latter value, reported again in Figs. ~\ref{fig_15}, ~\ref{fig_16} and ~\ref{fig_17}, show good agreement between model and data. In particular, the transmissions cut-on and cut-off frequencies coincide now with those measured experimentally. The performance parameters within the NIKA bands are summarised in the data in Fig.~ \ref{tab_2}. The transmissions are greater than 90\% in both bands, there is a clear improvement of the 45$^\circ$ cross-polarization and differential phase-shift on the 2-mm band but a deterioration in the 1-mm band. We note that the averaged cross-pol is below the -15 dB requirement across a 92\% bandwidth (105-285 GHz).

Although the performance of this device was not meant to meet the requirements, the main achievement of this R\&D phase was to understand which parameters could change during the various manufacturing processes. We have either evaluated or directly measured these parameters and assessed their variability range. We have built models with fitted parameters providing results very close to the experimental data, both for the single stacks and for the final assembled device. We conclude that, within the various parameters, the increase of the substrate refractive index and the change in the grid geometries play the major roles in the performance deviations of the manufactured device.

\section{NIKA2 mesh-HWP (M-HWP$_C$)}\label{sec6}
\begin{figure}[!ht]
\centering
\includegraphics[width=\columnwidth]{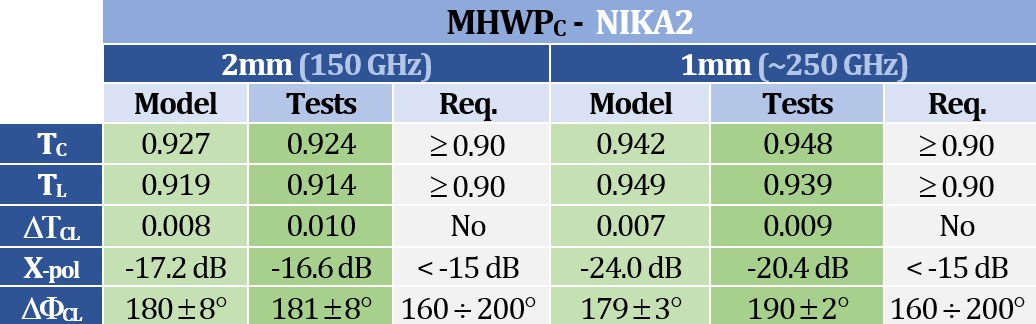}
\caption{Originally modelled, measured and required performance parameters of the M-HWP$_C$ across the two NIKA2 bands. The labels and colours are the same as described in the data in Fig.~ \ref{tab_1}. Note that this was the only plate designed to work across both bands.}
\label{tab_3}
\end{figure}

The results achieved through the development of the first two waveplates proved our ability to model the performance of the final devices with good accuracy. The NIKA2 requirements were similar to the NIKA ones (T$_C$ $\geq$ 0.90, T$_L$ $\geq$ 0.90 and $\Delta \Phi_{C-L}$ = 180$^{\circ}$ $\pm$ 20$^\circ$) but extended now to both bands. Some further improvement was required in terms of differential transmission because, in the previous realizations, we had a slight unbalance between the transmissions along the axes in the 2-mm band (see Figs.~ \ref{tab_1} and \ref{tab_2}). As mentioned, this was due to the larger losses introduced by the narrow inductive grids near their low frequency cut-offs, around 100 GHz. A solution was to move the cut-off frequency to lower frequencies. A way to achieve this was to relax some of the parameter constraints in the optimisation procedure. Given the narrow width of the two NIKA2 bands there was no need to achieve high performance across the whole 110-320 GHz frequency range. The optimisation procedure was then modified to maximise the transmissions and minimise the phase error only within the NIKA2 bands. This resulted in a dip in the inductive axis transmission around 200 GHz which had no impact on the overall performance because this region would be filtered out. The differential transmission was eventually kept within 1\% in both bands (see Fig.~ \ref{tab_3}). The 45$^\circ$ cross-polarization had targeted minima within each of the bands, so forcing the differential phase-shift to cross the ideal 180$^\circ$ within them. 

We note that at the end of each design optimisation the output grid parameters need to be rounded to values which allow periodicity between grids. This is needed for their finite-element computation which is based on periodic boundaries. The overall effect is a small deviation from the previously optimised performance. 

\begin{figure}[!ht]
\centering
\includegraphics[width=\columnwidth]{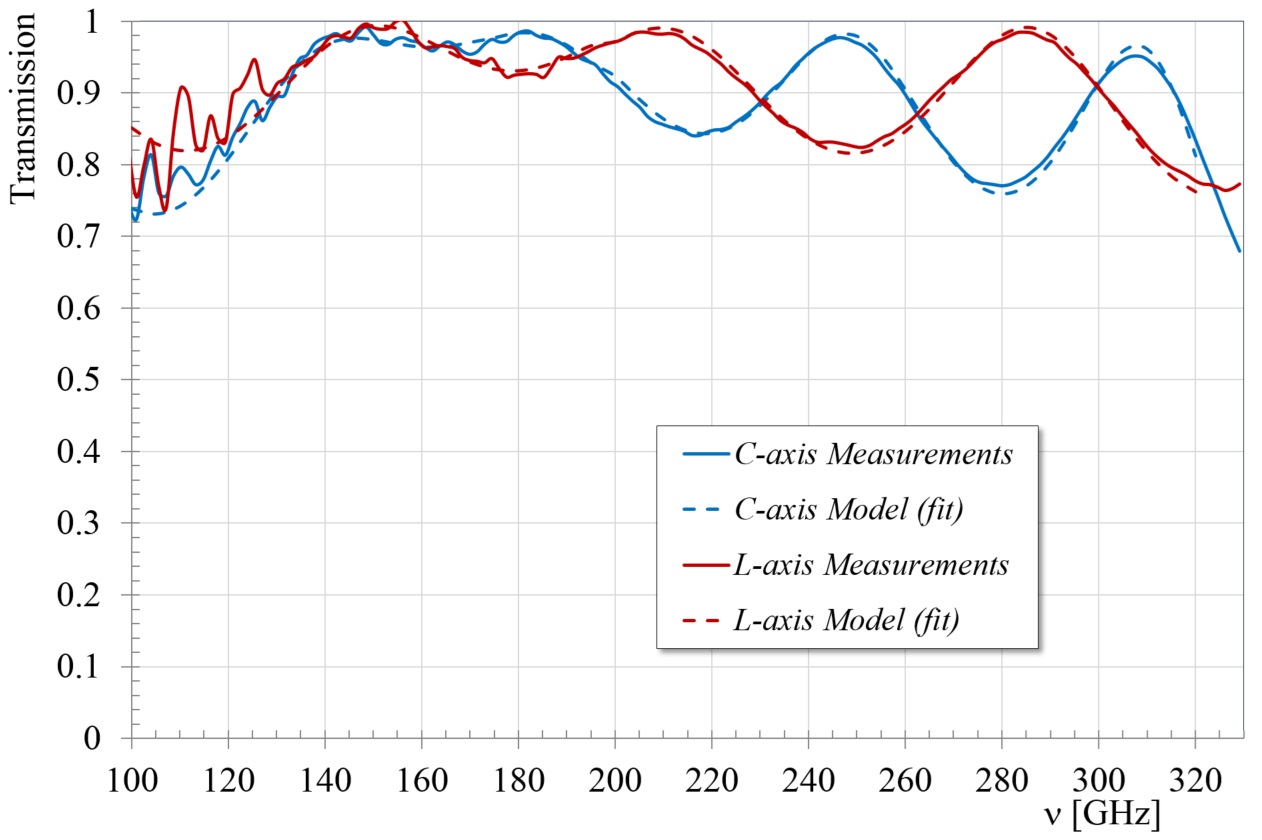}
\caption{M-HWP$_C$: Capacitive stack fitted model and measured transmissions along the reactive and transparent axes.}
\label{fig_18}
\end{figure}

\begin{figure}[!ht]
\centering
\includegraphics[width=\columnwidth]{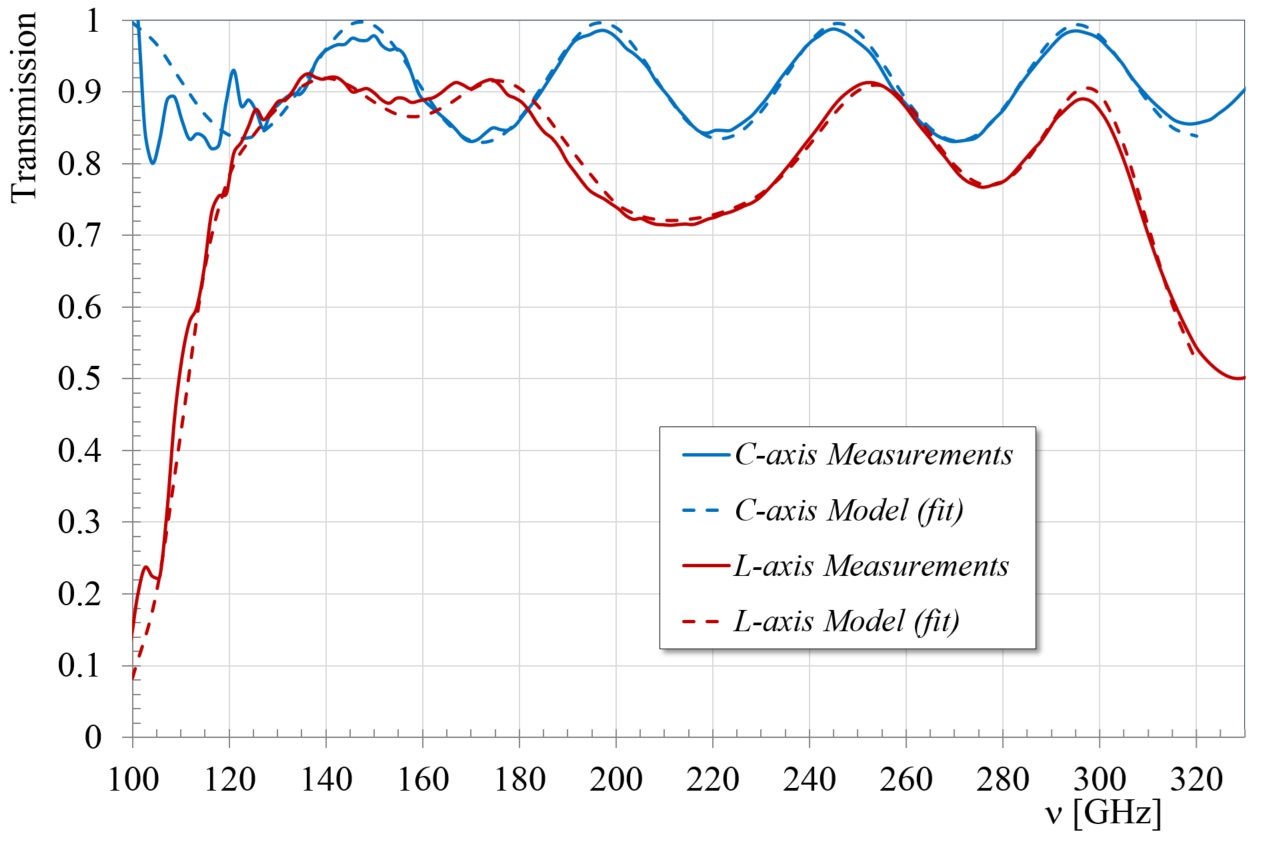}
\caption{M-HWP$_C$: Inductive stack fitted model and measured transmissions along the reactive and transparent axes.}
\label{fig_19}
\end{figure}

The grids were manufactured using the processes described above, this time more accurately controlling the grid etching processes using a weaker etch solution and microscopic imaging of the grids. The widths of inductive lines were chosen to be at least 20 $\mu$m wide in order to reduce their losses and to minimise the over-etching effects on the performance faced earlier. The capacitive and inductive stacks (6-C and 6-L) were assembled and pressed separately and, as previously, were slightly thicker than expected, within the percent level. The transmission of both stacks was measured with the FTS along their reactive and transparent axes, as reported in Fig.~\ref{fig_18} and Fig.~\ref{fig_19}.

For each stack, a parameter optimisation was run in order to determine the PP refractive index and the copper conductivity. The optimisation was run by minimising the difference between the modelled and the measured transmissions in a discrete number of frequency points. It was found that the stacks had slightly different refractive indices, \textit{n}$_{PP-C}$=1.517 and \textit{n}$_{PP-L}$=1.535, and conductivity $\sigma_{Cu}$ $\simeq$  3.00e7 Sm$^{-1}$. The over-etching in the C-stack ranged within 1-6 $\mu$m, while in the L-stack we actually observed 2-3 $\mu$m under-etching of the lines. As shown in Fig.~\ref{fig_18} and Fig.~\ref{fig_19}, the above values provided very good agreement between data and models in all four transmission curves.

\begin{figure}[!ht]
\centering
\includegraphics[width=\columnwidth]{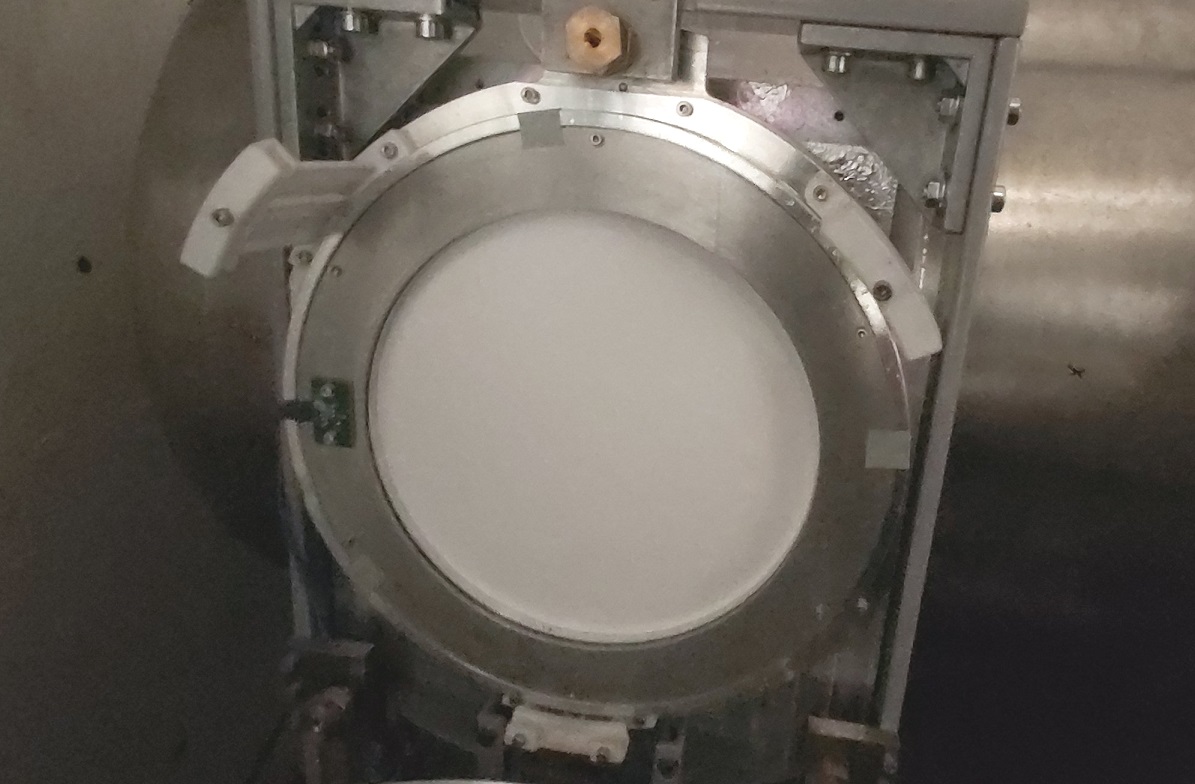}
\caption{M-HWP$_C$ device mounted in front of the NIKA2 istrument.}
\label{fig_20}
\end{figure}

\begin{figure}[!ht]
\centering
\includegraphics[width=\columnwidth]{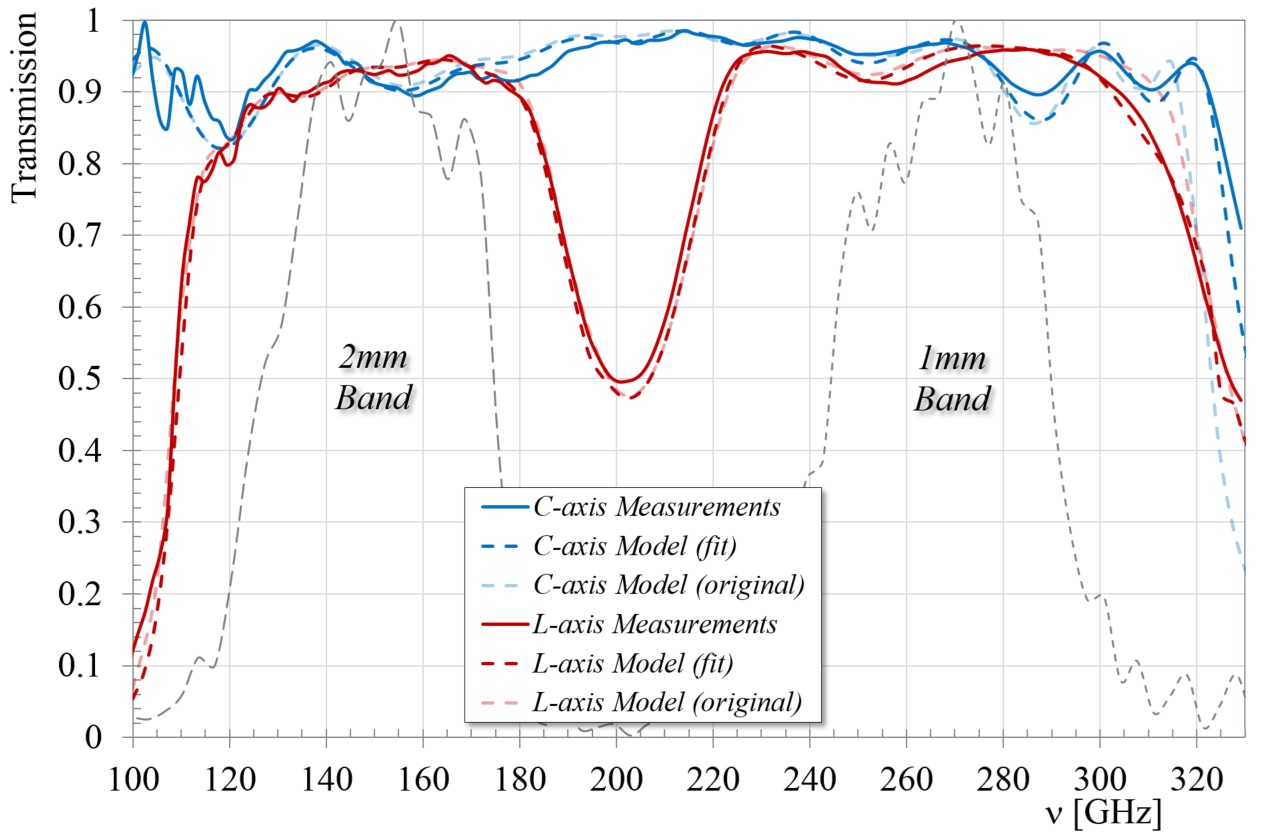}
\caption{M-HWP$_C$: Capacitive and inductive axes original/fit model and measured transmissions.}
\label{fig_21}
\end{figure}

\begin{figure}[!ht]
\centering
\includegraphics[width=\columnwidth]{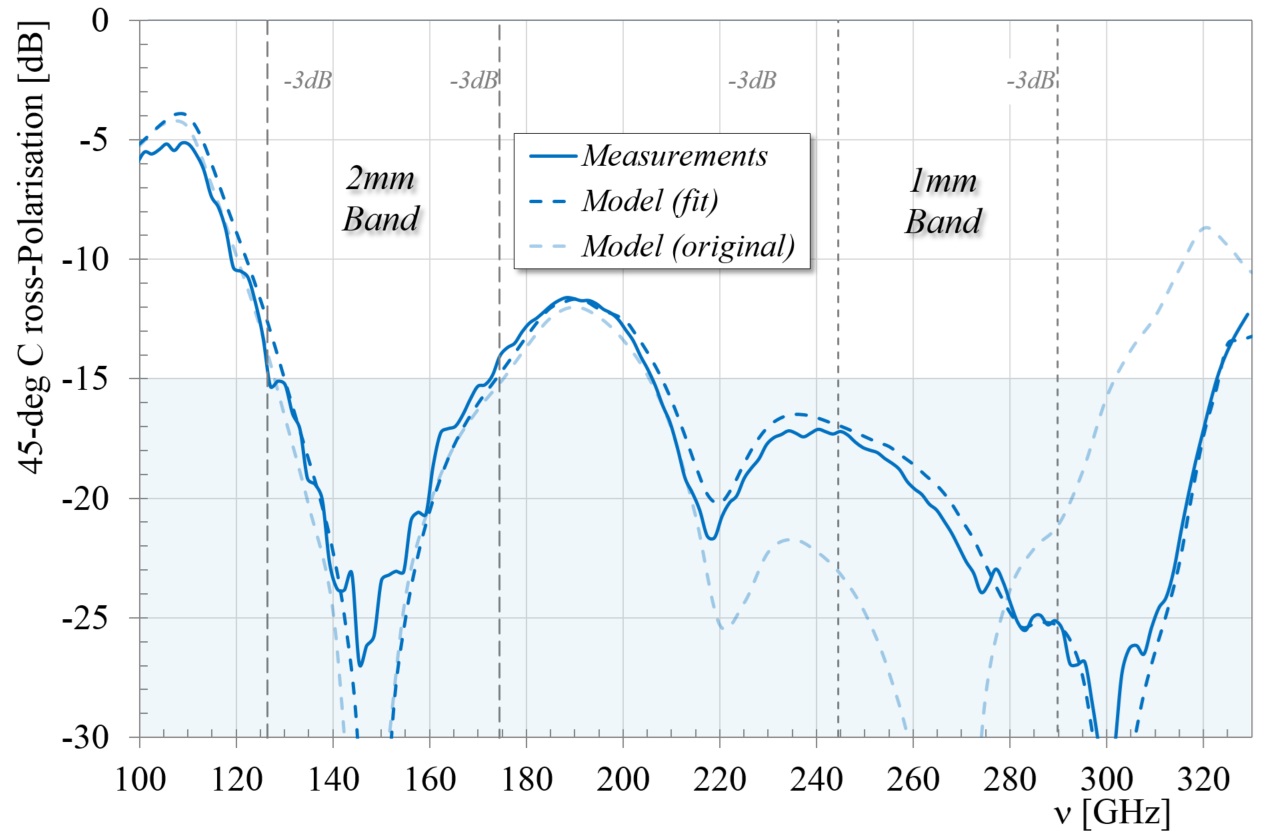}
\caption{M-HWP$_C$: 45$^\circ$ cross-polarization original/fit model and measurements.}
\label{fig_22}
\end{figure}

The stacks of the final M-HWP$_C$ were eventually bonded and AR-coated as before in a final hot-pressing process. A picture of the device, mounted into the NIKA2 instrument, is shown in Fig.~\ref{fig_20}. The device had an overall thickness of ~3.9 mm. FTS measurements of the axes transmissions and 45$^\circ$ cross-polarization are shown in Fig.~\ref{fig_21} and Fig.~\ref{fig_22}. A final fine-tuning optimization was run to take into account the parameters that could still change in the final process, i.e. the thickness and the refractive indices of each stacks. This led to a final increase of the stacks' refractive indices: \textit{n}$_{PP-C}$=1.524 and \textit{n}$_{PP-C}$=1.556, the latter being similar to that which was assumed in the design (\textit{n}$_{PP}$ =1.55), following the results of the M-HWP$_B$ development. A small increase in the stacks' thicknesses was observed, although still below 1\% from the nominal values. 

\begin{figure}[!ht]
\centering
\includegraphics[width=\columnwidth]{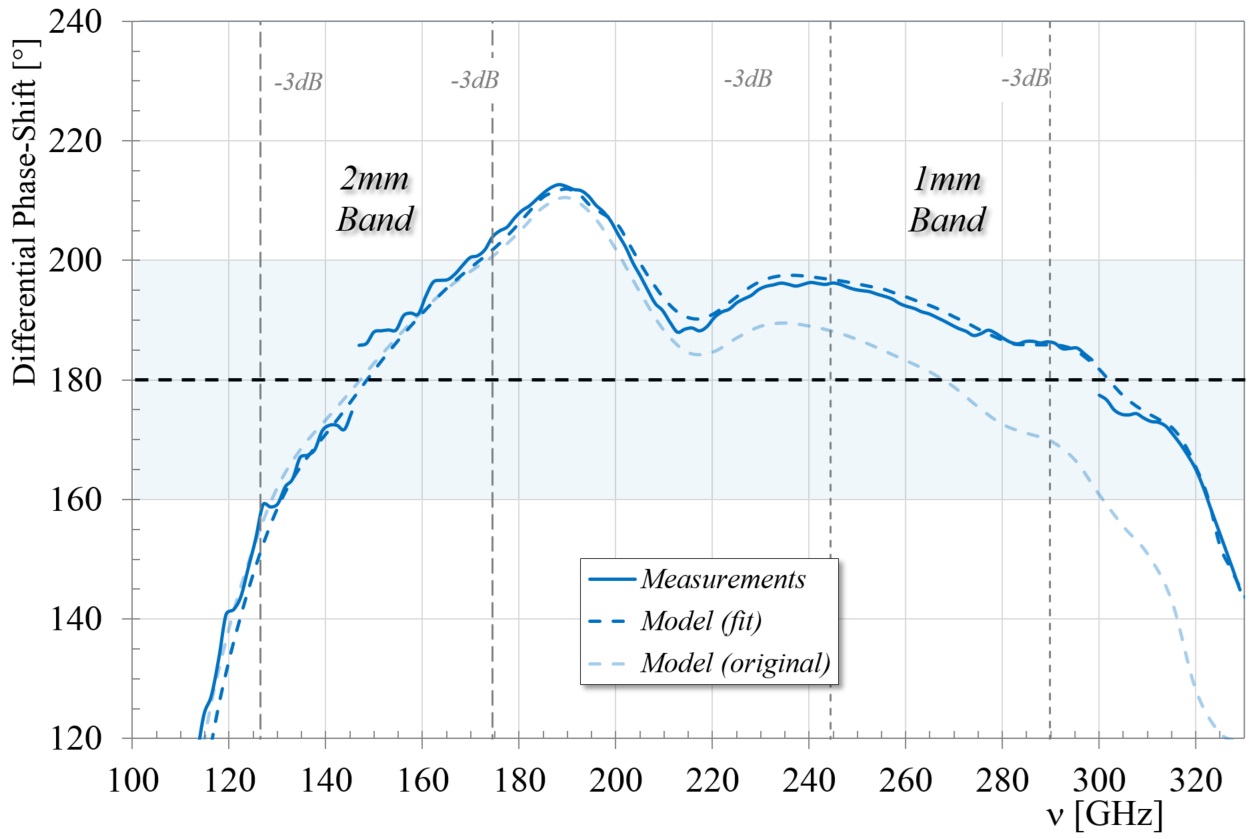}
\caption{M-HWP$_C$: Differential phase-shift original/fit model and measurements.}
\label{fig_23}
\end{figure}

The fitted-parameter models gave very good agreement with the measured data for both axes and 45$^\circ$ cross-polarization (Fig.~\ref{fig_21} and Fig.~\ref{fig_22}). In the same figures we have also reported the expected performance of the original design, i.e. the model results with the nominal values. The differential phase-shift, extracted from the data as before, is reported in Fig.~\ref{fig_23} along with its fitted and original model. We note that the measured data is accurately reproduced by the fitted models in all the performance parameters. In addition, these curves are now very close to the original models with the nominal values. 

The M-HWP$_C$ transmissions and differential phase-shift, weighted by the band transmission profiles, are reported in Fig.\ref{tab_3}. The device met the 2- and 1-mm NIKA2 band requirements for all the parameters. We note that the differential transmission is now kept at 1\% level in both bands, meaning that the unbalance evident in the previous devices has been successfully addressed and solved. We note that, given the noise level in the NIKA2 band transmissions (see Fig.~\ref{fig_03}), we have integrated the above values between 109-182 GHz and 221-304 GHz respectively for the 2- and 1-mm bands, i.e. within the ranges where the band transmission profiles are above their respective noise level (~5\% and ~10\%).

\begin{figure}[!ht]
\centering
\includegraphics[width=\columnwidth]{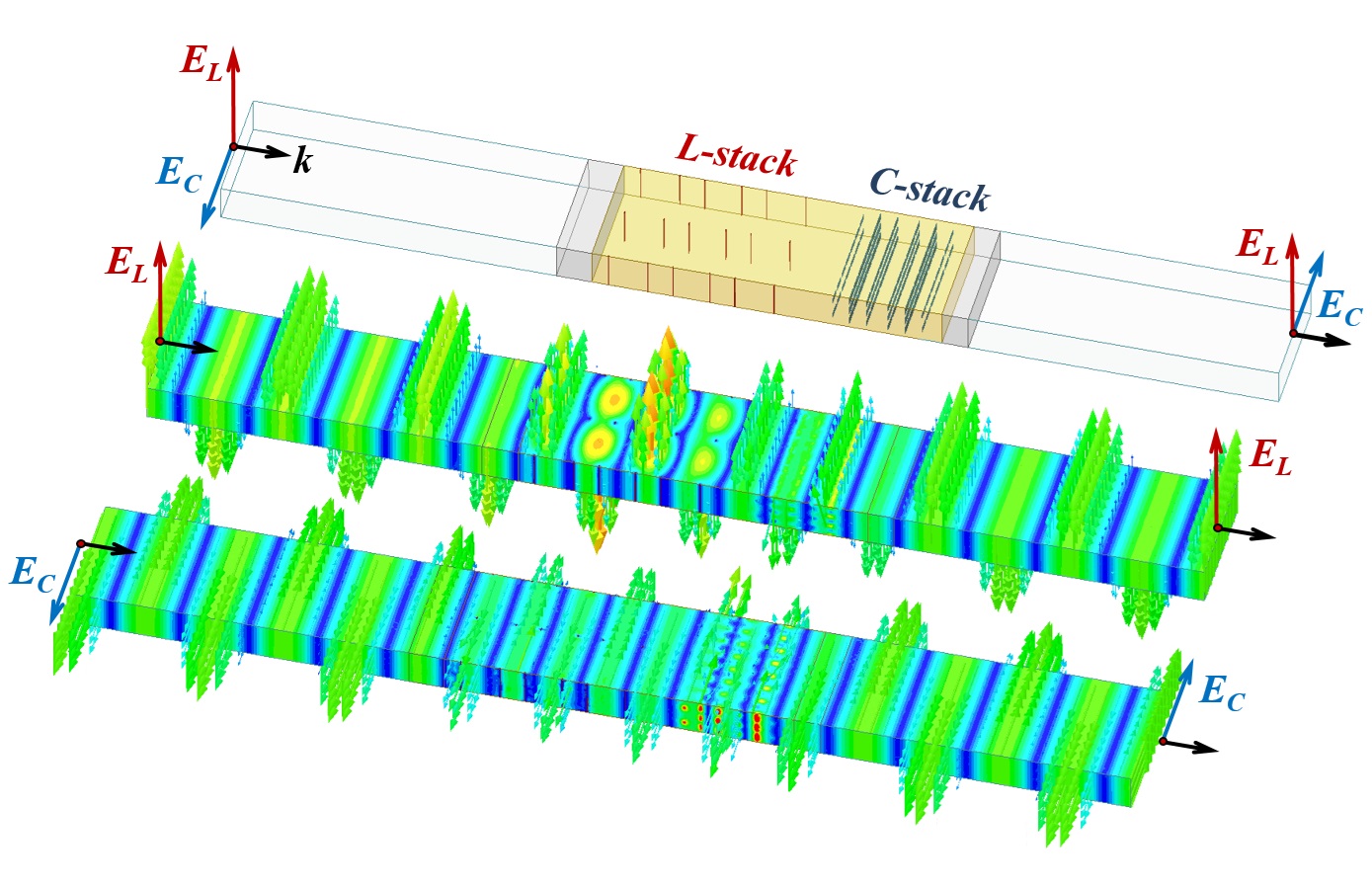}
\caption{M-HWP$_C$ finite-element unit cell model and simulations, run at 230 GHz, showing the propagation of the orthogonal electric fields, aligned with the L and C grids. Note that, at those specific positions, the outgoing vectors are in-phase (E$_L$) and out-of-phase (E$_C$) with respect to the incoming ones, so showing the expected introduction of a ~180$^\circ$ phase-shift between the two polarization components.}
\label{fig_24}
\end{figure}

To conclude, as a representative example of all the M-HWPs discussed before, in Fig.~\ref{fig_24} we report the results of a finite element simulation of M-HWP$_C$. The simulation, run at 230 GHz, shows the interaction of an electromagnetic wave going through the mesh HWP and its dependence on the polarization alignment with respect to the capacitive or inductive axis.

\section{Polarimetry with NIKA and NIKA2 instruments}\label{sec7}
Adopting the strategies described in Sec.~\ref{sec2}, we have performed polarization observations with both NIKA \citep{ritacco2017,ritacco2018} and NIKA2 \citep{ajeddig,ritacco2020} on compact and extended galactic and extra-galactic objects. In order to show the performance of the NIKA and NIKA2 polarimeters we recall here the maps obtained on a supernova remnant, the Crab nebula. An extended study by using the NIKA polarization results as well as all existing polarization data at submm and mm wavelengths is reported in \cite{ritacco2018}. We anticipate that, thanks to the larger instantaneous field-of-view, better sensitivity and optimised configuration, the NIKA2 mapping speed in polarization will be more than an order of magnitude higher when compared to NIKA.

\subsection{On sky observations}\label{subsec7.1}
In polarimetric mode the signal received by NIKA/NIKA2 detectors is a combination of the three Stokes parameters I, Q, U; being the polarized signal in Q, U modulated at four times the mechanical rotation frequency $\omega$ of the HWP. 
In order to have the right compromise between the sampling rate, the telescope speed and to limit systematic effects the mechanical rotation frequency $\omega$ for both instruments is set to $\sim$ 2.98 Hz \citep{ritacco2017}.
The polarization data are calibrated using standard calibration procedures developed for total intensity observations \citep{2019Perotto}. In addition, in order to treat polarization data three major steps in the data analysis have to be considered: i) the HWP Synchronous Signal subtraction that appears as a spurious signal modulated at harmonics of $\omega$; ii) the instrumental polarization characterization that shows up as a bipolar (NIKA) and quadrupolar pattern (NIKA2); iii) demodulation procedure to produce independent Stokes I, Q, U timelines to project onto final maps.
\cite{ritacco2017} present the data analysis pipeline specifically developed to treat these effects for NIKA, extended then to NIKA2. The commissioning of the NIKA2 polarimeter started in fall 2017 and it is still ongoing due to few technical issues. The first lights obtained are shown in \cite{ajeddig, ritacco2020}.

\subsection{NIKA/NIKA2 polarimeters on the sky}
\begin{figure}[!ht]
\centering
\includegraphics[width=0.45\columnwidth]{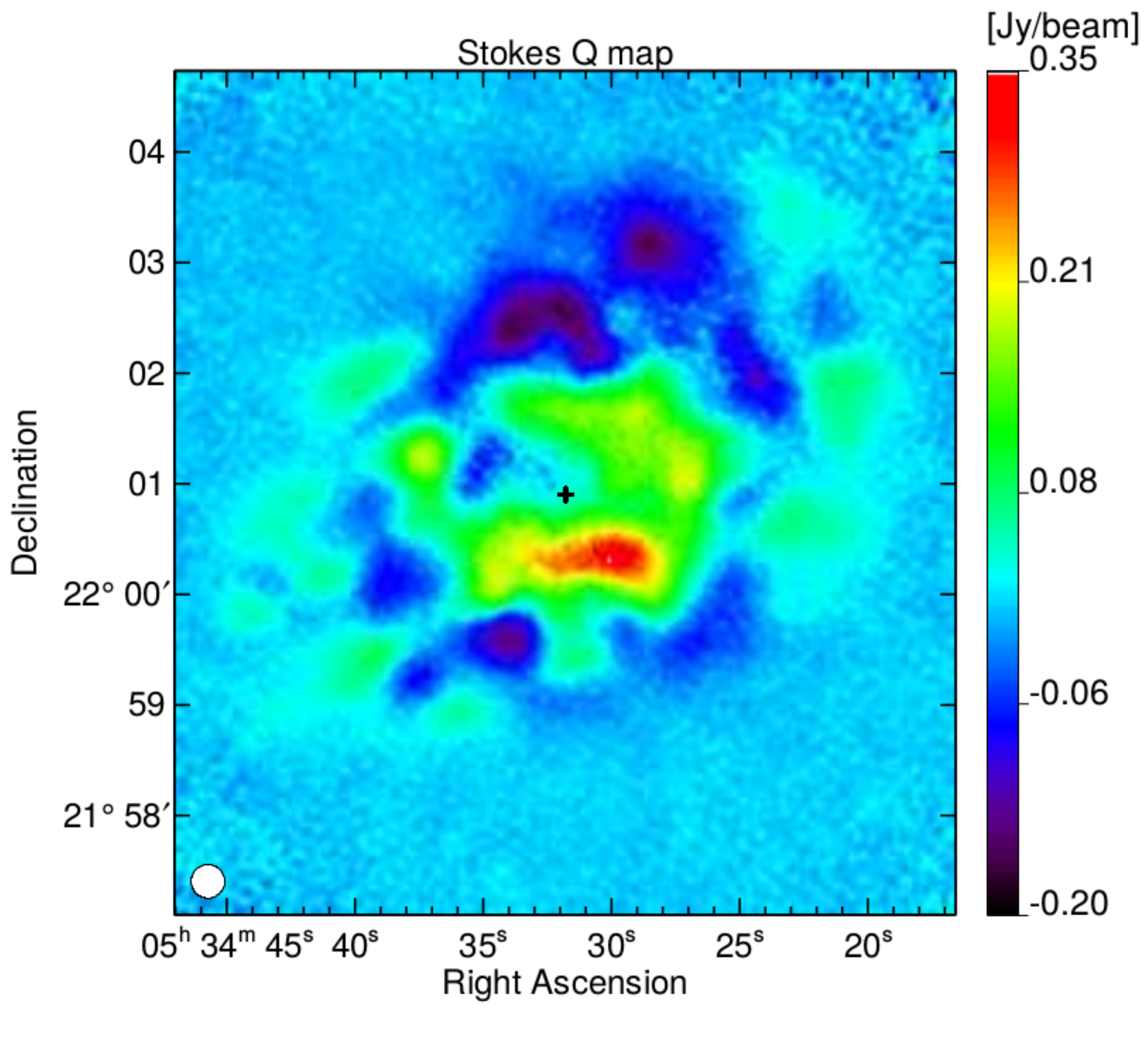}
\includegraphics[width=0.45\columnwidth]{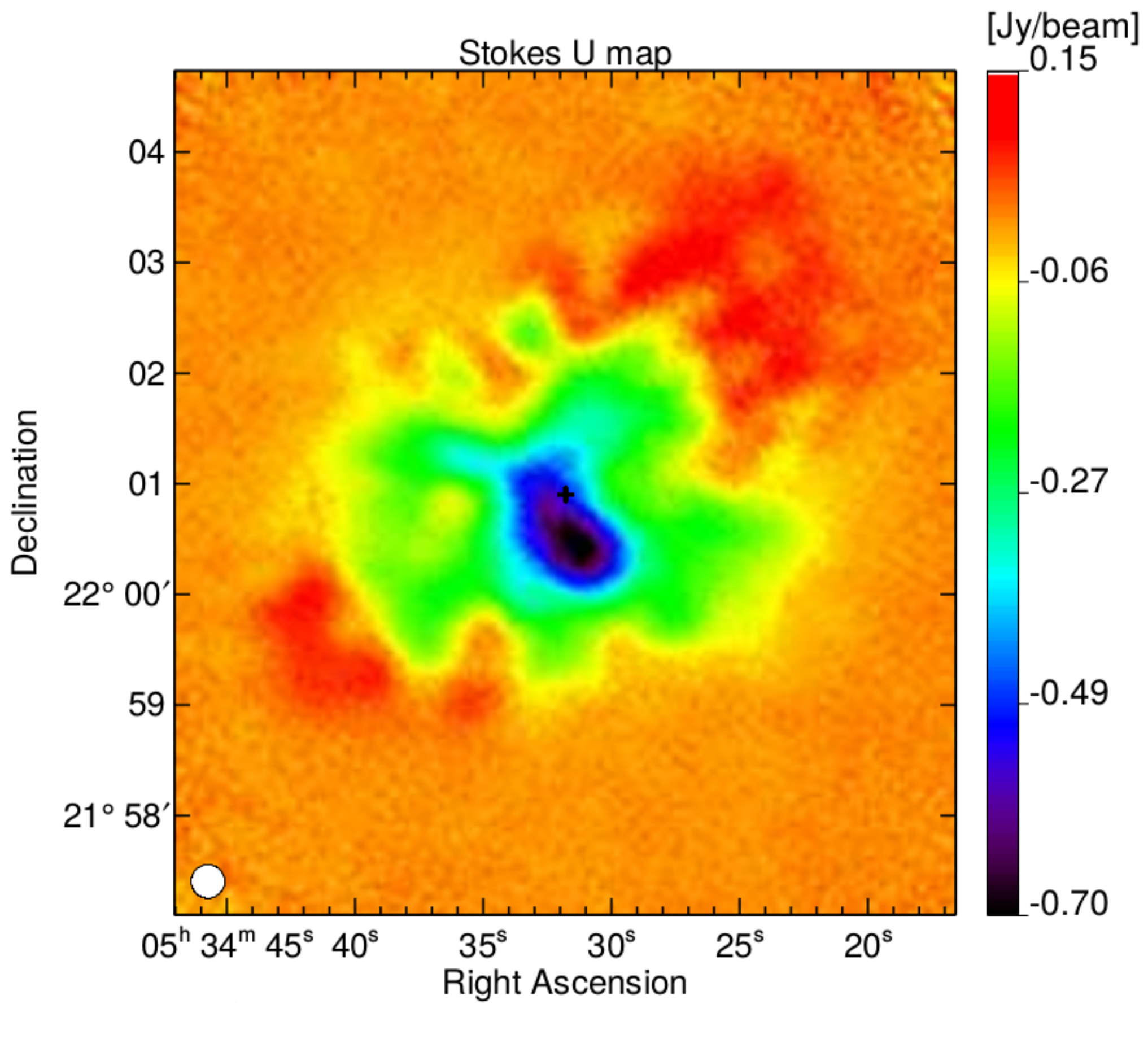}
\includegraphics[width=0.45\columnwidth]{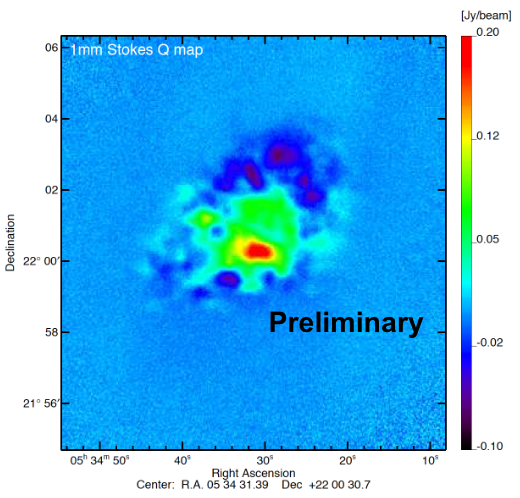}
\includegraphics[width=0.45\columnwidth]{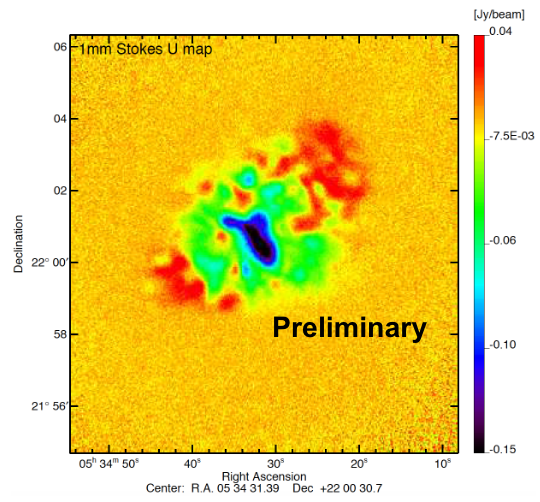}
\caption{From left to right Stokes Q and U maps obtained with NIKA at 150 GHz (top) \cite{ritacco2018} and NIKA2 at 260 GHz (bottom) \citep{ritacco2020}.}
\label{fig_25}
\end{figure}

As an illustration we summarize here the results obtained on the Crab nebula with NIKA and NIKA2 polarimeters.
The Crab nebula is a supernova remnant that exhibits a synchrotron emission in a large range of frequency, from radio to X-ray. \cite{ritacco2018} demonstrated that the spectral index in polarization at submm and mm wavelengths is consistent with total intensity spectral index. We therefore expect to have the same polarization in the two mm bands of interest. Fig.~\ref{fig_25} shows the polarization maps in terms of Stokes Q and U obtained with NIKA at 150 GHz \citep{ritacco2018} and NIKA2 at 260 GHz \citep{ritacco2020}. Although the maps obtained with NIKA2 are only preliminary we can already notice the high consistency of the polarized signal observed on both maps between NIKA and NIKA2. Notice that for more or less the same time of observation it was not possible to get a sufficient SNR on the maps at 1mm with NIKA. 
A more detailed study on this comparison is beyond of the scope of this paper and will be reported by the end of the NIKA2 polarimeter commissioning.
 
\section{Conclusions}\label{sec:conclusions}
The different challenges of modern astrophysics in the polarization field require high precision and control of systematics, which is possible with HWPs devices like the one used in NIKA and NIKA2.  
The NIKA and NIKA2 instruments are dual-band imaging instruments observing simultaneously in the 1mm and 2mm atmospheric windows. Polarization modulation can be implemented by using a single rotating HWP in front of these instruments. The HWP is required to have a large fractional bandwidth, of the order of ~3:1. 

For this purpose we have pushed further the development of half-wave plates based on the mesh-filter technology. Three different mesh-HWPs were designed, manufactured and tested. The first device, while showing some variations from the expected performance, met the NIKA single-band requirements. The second device was developed to understand, isolate and control the variations of the geometrical and physical parameters occuring during the manufacture of the waveplates. The knowledge acquired in developing the first two devices was successfully implemented in the third device, which had more challenging dual-band requirements set by the NIKA2 instrument.

The first mesh-HWP was installed in the NIKA instrument at IRAM providing high precision simultaneous observations at 150 GHz and 260 GHz of the polarization in different targets, from quasars to galaxies and star forming regions \citep{ritacco2017}. It has also provided the first polarization observation of the Crab nebula at 150 GHz with an angular resolution (FWHM $\sim$18 arcsec) \citep{ritacco2018}, as well as presenting for the first time an in-depth study of the spectral dependence of its polarization at millimeter wavelenghts.

The third mesh-HWP was installed in the NIKA2 instrument and tested along the last two years while NIKA2 was already used for open-time observations in total intensity. Despite the expectations, the commissioning of the NIKA2 polarization system is taking longer than expected because of a number of technical challenges and adverse weather conditions. However, the first light results are already promising. 

We note that mesh-HWPs, like mesh-filters, are very light and robust devices. Their anti-reflection coating, which is part of their structure, does not suffer delamination at cryogenic temperatures, as in Sapphire-based wavaplates. In addition, a Sapphire HWP operating across the NIKA2 bands would require a 3-plate Pancharatnam design with a total thickness of $\sim$6.3 mm and  a weight of $\sim$780 g (assuming a diameter of 200 mm). The NIKA2 mesh-HWP is 3.9 mm thick and has a weight of $\sim$120 g, a factor 6.5 lighter.  The final advantage of a large bandwidth mesh-HWP is the fact that the waveplate axes are uniquely defined by the metal grids and their orientation do not change with the frequency, as it happens in multiplate Pancharatnam devices.

\begin{acknowledgements}
We would like to thank the IRAM staff for their support during the campaigns. The NIKA and NIKA2 dilution cryostats have been designed and built at the Institut N\'eel. We acknowledge the crucial contribution of the Cryogenics and Electronics Groups in Grenoble. This work has been partially funded by the Foundation Nanoscience Grenoble and the LabEx FOCUS ANR-11-LABX-0013. This work has been supported by the French National Research Agency under the contracts ``MKIDS'', ``NIKA'' and ANR-15-CE31-0017 and in the framework of the ``Investissements d'avenir'' program (ANR-15-IDEX-02). This work has benefited from the support of the European Research Council Advanced Grant ORISTARS under the European Union\'s Seventh Framework Programme (Grant Agreement no. 291294). 
\end{acknowledgements}


\bibliographystyle{aat}

\bibliography{biblio}


\end{document}